\documentclass[aps,prx,longbibliography,reprint]{revtex4-2}
\usepackage{graphicx}
\usepackage{amsmath}
\usepackage{amsfonts}
\usepackage{amssymb}
\usepackage{bbm}
\usepackage[]{units}
\usepackage{hyperref}
\usepackage{braket}
\usepackage{dsfont}
\usepackage{mathdots}
\usepackage{natbib}
\usepackage[T1]{fontenc}

\newcommand{\nnb}{\nonumber \\}

\newcommand{\bv}{\left( \begin{array}{c}}
\newcommand{\ev}{\end{array} \right)}

\newcommand{\st}[1]{_{\text{#1}}}
\newcommand{\bo}[1]{\boldsymbol{#1}}

\begin{document}
\title{Simple framework for systematic high-fidelity gate operations}
\author{Maximilian Rimbach-Russ}
\author{Stephan G.J. Philips}
\author{Xiao Xue}
\author{Lieven M.K. Vandersypen}
\affiliation{QuTech and Kavli Institute of Nanoscience, Delft University of Technology, Lorentzweg 1, 2628 CJ Delft, The Netherlands}
\begin{abstract}

Semiconductor spin qubits demonstrated single-qubit gates with fidelities up to $99.9\%$ benchmarked in the single-qubit subspace. However, tomographic characterizations reveals non-negligible crosstalk errors in a larger space. Additionally, it was long thought that the two-qubit gate performance is limited by charge noise which couples to the qubits via the exchange interaction. Here, we show that coherent error sources such as a limited bandwidth of the control signals, diabaticity errors, microwave crosstalk, and non-linear transfer functions can equally limit the fidelity. We report a simple theoretical framework for pulse optimization that relates erroneous dynamics to spectral concentration problems and allows for the reuse of existing signal shaping methods on a larger set of gate operations. We apply this framework to common gate operations for spin qubits and show that simple pulse shaping techniques can significantly improve the performance of these gate operations in the presence of such coherent error sources. The methods presented in the paper were used to demonstrate two-qubit gate fidelities with $F>99.5\%$ in Ref.~[Xue et al, Nature \textbf{601}, 343]. We also find that single and two-qubit gates can be optimized using the same pulse shape. We use analytic derivations and numerical simulations to arrive at predicted gate fidelities greater than 99.9\% with duration less than $4/(\Delta f)$ where $\Delta f$ is the difference in qubit frequencies.
\end{abstract}
\maketitle
%%%%%%%%%%%%%%%%%%%%%%%%%%%%%%%%%%%%%%%%
\section{Introduction}
\vspace{-10pt}
Spin qubits based on electrons confined in quantum dots (QDs)~\cite{hansonSpinsFewelectronQuantum2007} are a leading candidate for long-term applications in quantum information processing. They provide long relaxation times~\cite{morelloSingleshotReadoutElectron2010,mauneCoherentSinglettripletOscillations2012,plaSingleatomElectronSpin2012,yangOrbitalValleyState2012,kawakamiElectricalControlLonglived2014,muhonenStoringQuantumInformation2014,veldhorstAddressableQuantumDot2014,engIsotopicallyEnhancedTriplequantumdot2015,reedReducedSensitivityCharge2016,lawrieSpinRelaxationBenchmarks2020,burkardSemiconductorSpinQubits2021} and their lithographic fabrications allow for dense and scalable qubit architectures~\cite{vandersypenInterfacingSpinQubits2017,veldhorstSiliconCMOSArchitecture2017}. Using isotopically enriched silicon (Si)~\cite{zwanenburgSiliconQuantumElectronics2013} or germanium (Ge)~\cite{scappucciGermaniumQuantumInformation2020} in favor of gallium arsenide (GaAs)~\cite{hansonSpinsFewelectronQuantum2007} as the host material for the quantum dots allows for significant longer decoherence times due to the low abundance of nuclear spins. One common feature of all spin qubits is the need for electric control on the nanoscale which typically also couples the system to electrical noise.  

Depending on the host material single-qubit gates are either implemented using electron spin resonance (ESR)~\cite{koppensDrivenCoherentOscillations2006,dehollainOptimizationSolidstateElectron2016,veldhorstTwoqubitLogicGate2015} or electric-dipole spin resonance (EDSR)~\cite{nowackCoherentControlSingle2007a,yonedaQuantumdotSpinQubit2018,zajacResonantlyDrivenCNOT2018,watsonProgrammableTwoqubitQuantum2018,hendrickxFastTwoqubitLogic2020,hendrickxFourqubitGermaniumQuantum2021} by applying microwave signals at the qubit resonance frequency.

On the other hand, all-electrical two-qubit gates can be implemented using dc gate voltage pulses that switch on and off the exchange interaction~\cite{lossQuantumComputationQuantum1998}. However, the originally proposed universal $\sqrt{\textsc{swap}}$ gate~\cite{lossQuantumComputationQuantum1998} was found to be unpractical to yield high fidelities in the case finite differences in qubit frequencies~\cite{burkardPhysicalOptimizationQuantum1999,meunierEfficientControlledphaseGate2011} are introduced for qubit addressability~\cite{vandersypenInterfacingSpinQubits2017} and strong dephasing from charge noise in the regime $\unit[100]{MHz}<J<\unit[10]{GHz}$~\cite{pettaCoherentManipulationCoupled2005,dialChargeNoiseSpectroscopy2013,reedReducedSensitivityCharge2016}. In the presence of a finite difference in qubit frequencies, the adiabatic \textsc{cz} gate, where a conditional phase difference is acquired by an adiabatic exchange pulse, is less stringent to hardware at the cost of longer gate times~\cite{meunierEfficientControlledphaseGate2011}. Two-qubit gates with fidelities $F>99\%$~\cite{xueQuantumLogicSpin2022a,madzikPrecisionTomographyThreequbit2022a,noiriFastUniversalQuantum2022a,millsTwoqubitSiliconQuantum2022} were recently reported with highest fidelities achieved by the adiabatic \textsc{cz} gate~\cite{xueQuantumLogicSpin2022a,millsTwoqubitSiliconQuantum2022}. 

Even without the presence of decoherence, qubit operations can be subject to errors. These coherent errors can arise from miss-calibration, crosstalk, non-adiabicity, finite bandwidths, filtered signals, non-linear transfer functions, from certain approximations made such as the rotating wave approximation, and many other spectator and control errors~\cite{vandersypenInterfacingSpinQubits2017}. Depending on the specifics, coherent errors can easily be larger than those from decoherence. 

The standard approach for mitigating these errors is summarized in optimal control theory~\cite{glaserTrainingSchrodingerCat2015} which can be divided into three main approaches. Firstly, a geometric approach that rewrites the time evolution into Pontryagin’s Maximum Principle~\cite{pontryaginMathematicalTheoryOptimal1987}. The optimal control pulse is then given by the extreme conditions that satisfy the given boundary conditions. However, analytical solutions are mostly limited to small and simple systems. Secondly, fully numerical techniques, such as the GRAPE~\cite{khanejaOptimalControlCoupled2005} and CRAB algorithms~\cite{canevaChoppedRandombasisQuantum2011}, can be used to find a (hopefully) global minima of the error by varying parameters of the input signal. This comes at the cost of speed and flexibility to small modifications. Lastly, inherent error mitigation can be achieved via (enforced) adiabatic dynamics~\cite{bergmannCoherentPopulationTransfer1998,motzoiSimplePulsesElimination2009,ribeiroSystematicMagnusBasedApproach2017,theisCounteractingSystemsDiabaticities2018}. 

In this paper we want to provide a simple framework to reduce coherent errors based on the adiabatic approach. We start in Section~\ref{sec:theoFrame} by introducing a framework which allows us to separate the desired dynamics which is the target gate from erroneous dynamics that yields gate errors. We also show how existing methods from literature~\cite{motzoiSimplePulsesElimination2009,ribeiroSystematicMagnusBasedApproach2017,theisCounteractingSystemsDiabaticities2018} are captured within this framework and can be reapplied to a larger set of gate operations. We then apply this framework in section~\ref{sec:applications} to derive optimized pulse shapes that reduce the errors on the most widly used single- and two-qubit gates for spin qubits. Subsequently in section~\ref{sec:results}, we numerically demonstrate the effectiveness of the derived pulse shapes in the presence of incoherent noise sources and benchmark them via the average gate fidelity~\footnote{Throughout the paper fidelity always refer to the average gate fidelity.}. By significantly reducing the magnitude of coherent errors our simulations show that gate fidelities $F>99.9\%$ with duration less than $\unit[50]{ns}$ are feasible.

\begin{figure*}
    \centering
    \includegraphics[width=1.\textwidth]{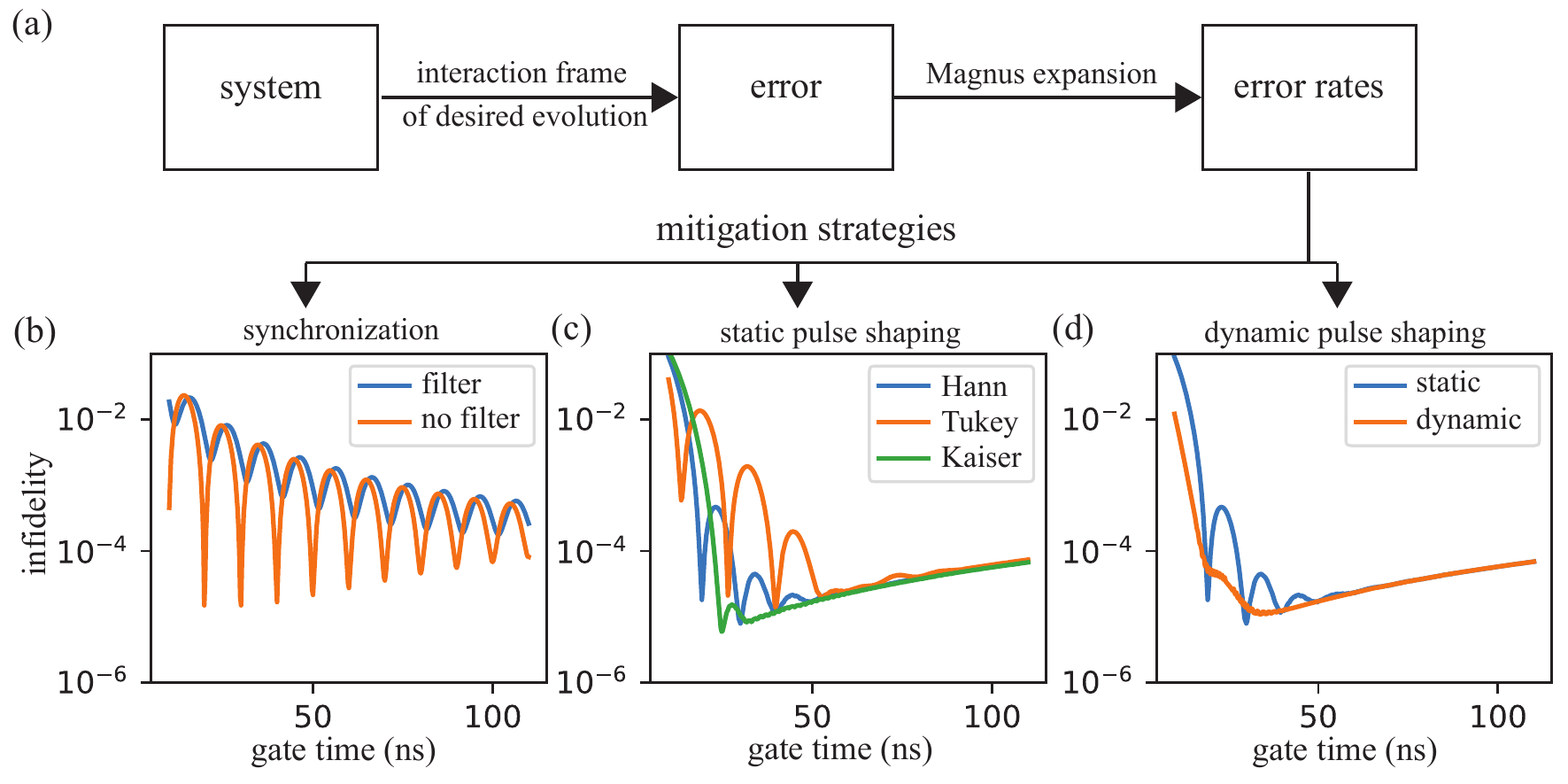}
    \caption{(a) Flowchart of the proposed framework for error mitigation. The system Hamiltonian $H=H_\text{ideal} + H_\text{rest}$ is separated into an ideal part covering the dominant interaction and a small erroneous part consisting of imperfections. After transforming into the interaction frame with respect to $H_\text{ideal}$ a Magnus expansion is performed to compute the error rates. (b)-(d) Mitigation strategies to reduce coherent errors shown for the example of a single-qubit $R_{x}(\pi/2)$ gate in a double-dot system with qubit frequency difference $\Delta E_z=\unit[100]{MHz}$. (b) Simulated infidelity of the gate operation as a function of the pulse-length with a control Hamiltonian that is instantanly turned on and off with (blue) and without (orange) filtering. Minima correspond to the synchronization condition~(see Refs.~\cite{russHighfidelityQuantumGates2018,heinzCrosstalkAnalysisSinglequbit2021}). (c) Simulated infidelity of the gate operation as a function of the pulse-length for different filtered pulse shapes optimized to concentrate the energy spectral density~\cite{martinisFastAdiabaticQubit2014}. (d) Simulated infidelity of the gate operation as a function of the pulse-length using a filtered Hann window with (blue) and without (orange) additional dynamic pulse shaping~\cite{motzoiSimplePulsesElimination2009}. Simulation parameters are discussed in Appendix~\ref{method:Simulation}. No incoherent noise sources are included. The gradual increase in infidelity for longer gate times is due to a residual exchange interaction between the qubits of $\unit[60]{kHz}$.}
    \label{fig:main_figure_1}
\end{figure*}

\section{Framework for optimizing pulse shapes}
\label{sec:theoFrame}
Fig.~\ref{fig:main_figure_1} displays our general framework for finding optimized pulse shapes to mitigate coherent errors.
We start considering a quantum system with a Hilbert space $\mathcal{H}$ which is described by a Hamiltonian $H$ of dimension $n\times n$. Ignoring any incoherent dynamics, the time-evolution from $t=0$ to a final time $t=t_g$ generated by a Hamiltonian is given by
\begin{align}
    U = \mathcal{T}\exp\left(-\frac{i}{\hbar}\int_0^{t_g} H(t^\prime) dt^\prime\right),
\end{align}
where $\mathcal{T}\exp$ is the usual time-ordering operator. We imply that $U$ describes an operation which is close to an ideal or a targeted operation described by the unitary operation $U_\text{ideal}$. We can now define the erroneous operation as 
\begin{align}
    \mathcal{E}=U^{\dagger}_\text{ideal}U .
\end{align}
The standard approach of estimating the errors is by measuring the average gate fidelity of the erroneous operation
\begin{align}
    F=\frac{|\text{tr}(\mathcal{E})|^2 +d}{d(d+1)},
    \label{eq:fidelity}
\end{align}
where $d$ is the dimension of the Hilbert space. For a noisy process described by a superoperator $\chi$ we replace $|\text{tr}(\mathcal{E})|^2\rightarrow \text{tr}(\chi)$ in Eq.~\eqref{eq:fidelity}. There are two standard approaches to experimentally access the gate fidelity, process tomography~\cite{nielsenQuantumComputationQuantum2000} and randomized benchmarking~\cite{magesanCharacterizingQuantumGates2012}, both requiring complex circuits and analysis and either susceptible to state preparation and measurement (SPAM) errors or limited in the information gain. However, much progress is made to increase the information gain and reduce the susceptibility to SPAM errors~\cite{aaronsonShadowTomographyQuantum2020,nielsenGateSetTomography2021}.

In order to estimate the errors, we separate $H=H_\text{ideal} + H_\text{rest}$ where $H_\text{ideal}$ is defined as generating the targeted gate
\begin{align}
U_\text{ideal} = \mathcal{T}\exp\left(-\frac{i}{\hbar}\int_0^{t_g} H_\text{ideal}(t^\prime) dt^\prime\right).
\label{eq:targetOperation}
\end{align}
This allows us the define an effective error Hamiltonian by switching into the interaction frame of the ideal gate operation~\cite{messiahQuantumMechanics1961,brinkmannIntroductionAverageHamiltonian2016,ribeiroSystematicMagnusBasedApproach2017} (see also Appendix~\ref{method:sparse_evolution})
\begin{align}
    H_\text{error}=U_\text{ideal}^\dagger(t)HU_\text{ideal}(t)-i\hbar U_\text{ideal}^\dagger(t)\dot{U}_\text{ideal}(t),
    \label{eq:errorHam}
\end{align}
where $\dot{U}$ is the derivative of $U$ with respect to time $t$. The effective error Hamiltonian now solely describes the erroneous operation and we can rewrite
\begin{align}
    \mathcal{E} &= \mathcal{T}\exp\left(-\frac{i}{\hbar}\int_0^{t_g} H_\text{error}(t^\prime) dt^\prime\right)\label{eq:Magnus_evolution}\\
    &= \exp\left(\sum_{n=1}^\infty \bar{H}_n\right)\label{eq:Magnus_exact}\\
    &\approx \exp\left(-\frac{i}{\hbar}\int_0^{t_g} H_\text{error}(t^\prime) dt^\prime\right).
    \label{eq:Magnus_approx}
\end{align}
From the first to the second line we applied the Magnus expansion with Magnus coefficients $\bar{H}_n$. We now make the standard assumptions such as small errors $||\mathcal{E}-\mathds{1}||\ll1$ (otherwise choose a closer desired gate) to ensure a fast converging Magnus series. In the last step we truncated the Magnus expansion at lowest order. Numerical simulations show that this order is sufficient to achieve quantum operations with errors $1-F<10^{-4}$.

As a next step we now compute and investigate transition rates caused by the erroneous operation (error rates) $|\text{tr}(\mathcal{E} O)|^2$ where $O$ is an operator of interest. The error rate $|\text{tr}(\mathcal{E} O)|^2$ is the probability to make a coherent error generated by $\mathcal{E}$ and ``measured'' by the operator $O$~\cite{blume-kohoutTaxonomySmallMarkovian2022}. From this point a general treatment is not encouraged since to completely optimize the gate operation, $(n^2-1)$ transition rates need to be optimized simultaneously. Fortunately, for most systems of interest the interaction is sparse, e.g., interaction is localized or limited to neighboring qubits. In many cases single-qubit operators $\sigma_x$, $\sigma_y$, $\sigma_z$ and two-qubit operators $\sigma_x\sigma_x$, $\sigma_y\sigma_y$, $\sigma_z\sigma_z$, where $\sigma_i$ is the $i=x,y,z$ Pauli-matrix, are sufficient. For spin qubits typical targeted gate operations are $e^{i\phi \sigma_z}$ gates due to different qubit frequencies~\cite{burkardSemiconductorSpinQubits2021}, $e^{i\phi \sigma_{x(y)}}$ single-qubit gates from resonant driving, and $e^{i \phi \sigma_z\sigma_z}$ two-qubit \textsc{cphase} gates. Under such sparse evolution the error rate typically takes the form (derivation see Appendix~\ref{method:sparse_evolution})
\begin{align}
    |\text{tr}(\mathcal{E} O)|^2 \approx \left|\frac{1}{\hbar} \int_0^{t_g} \sum_n g_n(t) e^{i f_n(t)/\hbar} dt \right|^2,
    \label{eq:sparseEvolution}
\end{align}
where $g_n(t)$ are functions connected with the erroneous evolution and $f_n(t)$ functions related to the desired evolution with each function depending on the choice of $O$. Without proof we now choose $O$ such that there is only a single term in the sum involved (this is indeed the case for our targeted applications) and perform the substitution~\cite{martinisFastAdiabaticQubit2014} 
\begin{align}
    \dot{f}(t)dt =\hbar \nu_f t_g ds
    \label{eq:substitution}
\end{align}
with a constant $\nu_f$. Under the implicit assumption $\dot{f}(t)\neq 0$ we arrive at the error rate
\begin{align}
    |\text{tr}(\mathcal{E} O)|^2 &\approx \left|\frac{1}{\hbar} \int_{s(0)}^{s(t_g)} \tilde{g}(t(s)) e^{i \nu_f t_g s} ds \right|^2, \label{eq:errorRate}\\
    &= S(\tilde{g}(\nu_f t_g)), \label{eq:errorESD}
\end{align}
with $\tilde{g}(t(s)) = g(t(s)) \frac{dt}{ds}(s)$ and the energy spectral density $S$. 
In the first line we approximated $\sin(x)\approx x$ and from the first to the second line we replaced the integral which corresponds to a short-time Fourier transformation with the energy spectral power of an input signal~\cite{martinisFastAdiabaticQubit2014}. As a consequence we have now shifted the task from minimizing the error rates to optimizing the energy spectral density, a task already investigated and in most cases solved. Conveniently, an experimental estimation of the error rates is much more efficient compared to the complex protocols required to compute the fidelity, since typically only a few expectation values are need to reconstruct the corresponding error rates.

\subsection*{Error mitigation strategies}
\subsubsection{Synchronization}
We know (without proof) that Eq.~\eqref{eq:errorESD} is an oscillatory function with respect to $\nu_f t_g$ due to the finite gate time $t_g$. This becomes clear by rewriting the short-time Fourier transformation as a convolution of the conventional Fourier transformation and the oscillating $\sin(x)/x$-function as the Fourier transform of the box function from the finite time interval $[0,t_g)$.
Synchronization is the concept of finding minima of Eq.~\eqref{eq:errorESD} which due to the oscillatory nature exist as shown in Fig.~\ref{fig:main_figure_1}~(b). The optimal pulse length $t_g$ or system parameter $\nu_f$ are then given by the minima of $S(\nu_f t_g)$. The concept of synchronization is best visualized in the special case of constant $\dot{f}(t)$ and $g(t)$ where even $S(\nu_f t_g)=0$ is possible. These minima correspond to cases where the undesired interaction "undoes" itself, e.g. phase evolution for a \textsc{swap} gate~\cite{burkardPhysicalOptimizationQuantum1999} or off-resonant Rabi oscillations~\cite{russHighfidelityQuantumGates2018,heinzCrosstalkAnalysisSinglequbit2021}. An advantage of this strategy is the absence of any complex pulse shaping. However, the requirement for simultaneous minima in the spectrum makes it difficult to scale beyond a handful of qubits~\cite{gullansProtocolResonantlyDriven2019,heinzCrosstalkAnalysisSinglequbit2021}. Additionally, filtering in the signal transmission greatly reduce the effectiveness of the gate operations as seen in Fig.~\ref{fig:main_figure_1}~(b).

\subsubsection{Static pulse shaping}
For purely real input signals, fast operations with consistently low error rates can be achieved using window functions $w(t)$ designed for optimized spectral concentration such as the discrete prolate spheroidal sequence (DPSS or Slepian), and the Dolph–Chebyshev window. Alternatively, if faster operations at the cost of larger coherent errors are desired, a Hamming window is the best choice. The optimized window functions typically require a high computation cost, high bandwidth, and high time-resolution thus restricting their use. For many applications with limited bandwidth and low-pass filters, equally small errors can be achieved using approximations such as the Kaiser window~\cite{kuoSystemAnalysisDigital1967}
\begin{align}
    w(t) = \mathcal{N} \mathcal{I}_{0}\left(\frac{2\lambda}{t_g} \sqrt{t(t_g-t)} \right)
\end{align}
with normalization constant $\mathcal{N}$ via $\int_0^{t_g}w(t)dt=t_g$ and $\mathcal{I}_{0}$ being the $0$-th order modified Bessel function. Alternatively, one can also use the
truncated Fourier expansions~\cite{martinisFastAdiabaticQubit2014} 
\begin{align}
    w(t) &= w_\text{even}(t) + w_\text{odd}(t)
    \label{eq:FourierShape}
\end{align}
with the even and odd decomposition
\begin{align}
w_\text{even}(t) &= \sum_{n=1}^{N}\lambda_{\text{even},n} \left[1-\cos\left(\frac{2\pi n t }{t_g}\right)\right],
\label{eq:FourierShapeEven}\\
w_\text{odd}(t) &= \sum_{n=1}^{N}\lambda_{\text{odd},n} \left[1-\sin\left(\frac{2\pi n t }{t_g}\right)\right].
\label{eq:FourierShapeOdd}
\end{align}
The coefficients $\left.\lambda_{\text{even}} = [1.0715,-0.0795,0.0043,0.0037]\right.$ and $\left.\lambda_{\text{odd}} = [0,0,0,0]\right.$ can be estimated from a Fourier decomposition of the slepian or from numerical minimization of the error rate~\cite{martinisFastAdiabaticQubit2014}. By using only even components a smooth pulse-shape is guarantied. A simple and popular yet effective pulse shape is the Hann window ($\left.\lambda_{\text{even}} = [1,0,0,0]\right.$) that corresponds to a cosine shape.
Another interesting pulse shape is the Tukey window defined as~\cite{xueQuantumLogicSpin2022a}
\begin{align}
    w(t,\lambda) = \begin{cases} 
      \frac{1}{2-\lambda}\left[1-\cos\left(\frac{2\pi t }{\lambda t_g}\right)\right] & 0 \leq t \leq \frac{\lambda t_g}{2} \\
      \frac{2}{2-\lambda} & \frac{\lambda t_g}{2} < t< t_g -\frac{\lambda t_g}{2} \\
       \frac{1}{2-\lambda}\left[1-\cos\left(\frac{2\pi (t_g-t) }{\lambda t_g}\right)\right] & t_g -\frac{\lambda t_g}{2}\leq t \leq t_g 
   \end{cases}
   \label{eq:Tukey_window}
\end{align}
which consists of two ramps $w_\text{even}(t)$ with $\lambda_{\text{even}}=[1,0,0,0]$ interleaved by a constant part. The Tukey pulse has the advantage of having a lower peak amplitude compared to the non-interleaved pulse at the cost of steeper flanks. For $\lambda=1$ it reduces to the Hann window. An interesting thought is also combining the Tukey window shape with the higher Fourier components which we will leave to the future. 

The optimal pulse design is then given by setting
\begin{align}
    \tilde{g}(s) = A w(t_g s)
\end{align}
with amplitude $A$ being estimated from the desired operation~\eqref{eq:targetOperation}. The relation between real time $t$ and dilated normal time $s$ is given by integrating Eq.~\eqref{eq:substitution} arriving at~\cite{martinisFastAdiabaticQubit2014}
\begin{align}
    t(s) = \int_{s(0)=0}^{s} \frac{\hbar \nu_f t_g }{\dot{f}(s^\prime)} ds^\prime
\end{align}
% with $\nu_f=\int_{0}^{1} \frac{\hbar t_g}{\dot{f}(s^\prime)} ds^\prime$. 
with $\hbar\nu_f t_g=f(t_g)-f(0)$. 
The inverted function is best acquired using numerical interpolation~\cite{martinisFastAdiabaticQubit2014}, e.g. \texttt{Mathematica} directly provides $\tilde{g}(t)$ using the command $\text{Interpolation}[\text{Table}[\lbrace t[s], \tilde{g}[s]\rbrace, \lbrace s, 0, 1\rbrace]]$ with sufficient sampling.
Fig.~\ref{fig:main_figure_1}~(c) shows the resulting infidelity of a gate operation for different pulse shapes. As expectec the Kaiser window outperforms the conventional cosine shape.

\subsubsection{Dynamic pulse shaping}
For resonantly driven gates, i.e., single-qubit gates for spin qubits~\cite{burkardSemiconductorSpinQubits2021}, the additional phase of the MW signal translates into complex $\tilde{g}(t(s))=\tilde{g}_R(t(s))+i \tilde{g}_I(t(s))$. Here, the real part $\tilde{g}_R(t(s))$ and imaginary part $\tilde{g}_I(t(s))$ correspond to the I/Q quadrature of the MW. This freedom can actively be used to further reduce the error rates compared to window functions~\cite{theisCounteractingSystemsDiabaticities2018}, e.g., the derivative by removal of adiabatic gate (DRAG)~\cite{motzoiSimplePulsesElimination2009,luceroReducedPhaseError2010} and Wah-Wah~\cite{schutjensSinglequbitGatesFrequencycrowded2013,vesterinenMitigatingInformationLeakage2014,theisSimultaneousGatesFrequencycrowded2016} protocols both allow to greatly suppress crosstalk from off-resonant drives. This can be visualized by integrating Eq.~\eqref{eq:errorRate} by parts with respect to the real part of the signal~\cite{motzoiOptimalControlMethods2011}
\begin{align}
    |\text{tr}(\mathcal{E}\,O|^2 &= S\left(\frac{\frac{d}{ds}\tilde{g}_R(t(s))}{i \nu_f t_g}+i \tilde{g}_I(t(s))\right)
\end{align}
with boundary conditions $\tilde{g}_R(0)=\tilde{g}_R(t_g)$. We get a complete cancellation of the error rate with
\begin{align}
    \frac{\frac{d}{ds}\tilde{g}_R(t(s))}{\nu_f t_g}= \tilde{g}_I(t(s)),
\end{align}
where the pulse shape of $\tilde{g}_R(t(s))$ can individually be optimized using window functions. Fig.~\ref{fig:main_figure_1}~(d) shows that using dynamic pulse shaping protocols (here DRAG) significantly reduces the error rate. For $\frac{dt}{ds}=\text{const}$ this exactly yields the DRAG condition $g_I\propto\dot{g}_R$. The advantage of this strategy is a very strong suppression of the error rate with the cost of additional power consumption which scales with the number of suppressed transitions~\cite{gambettaAnalyticControlMethods2011}. Our framework allows to generalize this powerful method to all systems with independent control over two orthogonal axes.

\section{Applications}
\label{sec:applications}
Before we apply our framework to optimize the performance of qubit operations we introduce a theoretical description of a spin qubit system.
The dynamics in the subsystem of two electron spins in the $(\cdots,1,1,\cdots)$ charge configuration of a multi-qubit network can be well-described by the Heisenberg model~\cite{lossQuantumComputationQuantum1998}
\begin{align}
H &=h\left( J\,(\boldsymbol{S}_1\cdot \boldsymbol{S}_2-\frac{1}{4}) +  \boldsymbol{B}_1\cdot\boldsymbol{S}_1+  \boldsymbol{B}_2\cdot\boldsymbol{S}_2\right)
\label{eq:ham_matrix}
\end{align}
with $\boldsymbol{S}_j = (\sigma_{x,j},\sigma_{y,j},\sigma_{z,j})^T/2$, where $\sigma_{k,j}$ is the Pauli-matrix acting on the spin in dot $j$ and $\boldsymbol{B}_j=(B_{x,j},B_{y,j},B_{z,j})^T$ being the magnetic field felt by the the electron in dot $j$.
For later convenience we define the average field $E_z=(B_{z,1}+B_{z,2})/2$ and the difference field $\Delta E_z= B_{z,2}- B_{z,1}$. Note that in this notation the magnetic field $\boldsymbol{B}$ and the exchange interaction $J$ are in units of $\unit{GHz}$.

Single-qubit gates are operated in the regime of negligible exchange interaction, $J\approx 0$, and implemented via resonant driving either via ESR or EDSR. Here, we do not distinguish between the two mechanisms and describe both by a static magnetic field $\boldsymbol{B}_{j,0}$ and an oscillating magnetic field $\boldsymbol{B}_{j,D}(t)$, such that $\boldsymbol{B}_j=\boldsymbol{B}_{j,0}+\boldsymbol{B}_{j,D}(t)$. We discuss here the case of two qubits driven by a single microwave tone. 

Exchange-based two-qubit gates are operated in the regime of $J> 0$.  For simplicity, we assume that all two-qubit gates are operated deep inside the (1,1) charge occupation regime at the symmetric operation point and electric control is maintained via virtual barrier gates~\cite{malinowskiSymmetricOperationResonant2017,reedReducedSensitivityCharge2016,xueQuantumLogicSpin2022a}. Theoretically, we describe the system using Hamiltonian~\eqref{eq:ham_matrix} and move into a rotating frame $R=\exp[-  i (2\pi\nu_D t+\theta) (S_{1}^z+S_{2}^z)]$ which leads in the standard basis $\lbrace\ket{\uparrow\uparrow},\ket{\uparrow\downarrow},\ket{\downarrow\uparrow},\ket{\downarrow\downarrow}\rbrace$ to the following Hamiltonian~\footnote{Here and below we use the notation that $\ket{\downarrow}$ and $\ket{\uparrow}$ correspond to ground and excited qubit state.}
\begin{widetext}
\begin{align}
    H_\text{RF}=\frac{h}{2}
\begin{pmatrix}
 2E_{z}-2\nu_D-\frac{\dot{\theta}}{2\pi} & B_{\perp,2}^* e^{-2\pi i \nu_D t} & B_{\perp,1}^* e^{-2\pi i \nu_D t} & 0\\
 B_{\perp,2} e^{2\pi i \nu_D t} & \Delta E_z - J & J & B_{\perp,1}^* e^{-2\pi i \nu_D t} \\
 B_{\perp,1} e^{2\pi i \nu_D t} & J & -\Delta E_z -J & B_{\perp,2}^* e^{-2\pi i \nu_D t} \\
 0 & B_{\perp,1} e^{2\pi i \nu_D t} & B_{\perp,2} e^{2\pi i \nu_D t} & -(2E_{z} - 2\nu_D-\frac{\dot{\theta}}{2\pi}) \\
\end{pmatrix},
\label{eq:ham_matrix_RF}
\end{align}
\end{widetext}
where $\nu_D$ is the drive frequency, $\theta$ the phase of the drive, and $B_{\perp,j}=B_{x,j}+i B_{y,j}$ the perpendicular component of the magnetic field felt by qubit $j=1,2$ with respect to the quantization axis.
%where $\tilde{E}_{\perp,2}^*=E_{\perp,2}^* e^{-2\pi i \nu_D t}$. 

\subsection{Resonant single-qubit gates}
Expected coherent errors are crosstalk and other spectator errors from off-resonant drives~\cite{philipsUniversalControlSixqubit2022}, non-linear driving giving rise to higher harmonics and phase shifts~\cite{undsethNonlinearResponseCrosstalk2022}, and frequency shifts from counter-rotating contributions (Bloch Siegert)~\cite{romhanyiSubharmonicTransitionsBlochSiegert2015}. To efficiently account for non-linear driving effects we expand the modulated (effective) magnetic field in terms of a Fourier series with respect to the drive frequency $\nu_{D}$ 
$
    \boldsymbol{B}_j(t) = \boldsymbol{\mathcal{B}}_{j,0}(t) + \sum_{k\neq 0} \boldsymbol{\mathcal{B}}_{j,k}(t) e^{-2\pi i \nu_{D} k t},
$
 where the Fourier components $\boldsymbol{\mathcal{B}}_{0}(t)$ and $\boldsymbol{\mathcal{B}}_{k}(t)$ are assumed to be slowly varying in the time interval $[0,\nu_{D}^{-1})$. 
 The Hamiltonian can be significantly simplified under the rotating wave approximation (RWA) where we keep stationary terms and disregard all terms which are modulated with frequency $k\nu_D$ with $|k|=1,2,3\cdots$. Corrections from violations of the rotating wave approximation (RWA) scale with $\Omega_\text{Rabi}/\nu_D\sim 10^{-3}$ thus are most of the times negligible for typical experimental conditions but can become important for ultra-fast gate operations~\cite{froningUltrafastHoleSpin2021,wangUltrafastCoherentControl2022} or driving at comparatively low frequencies~\cite{koppensDrivenCoherentOscillations2006}. In this section we focus on the RWA case and leave the latter to future investigations. We use the following Hamiltonian for our pulse optimization framework
 \begin{widetext}
 \begin{align}
    H_\text{RWA}=\frac{h}{2}
\begin{pmatrix}
 2(E_z -\nu_D)-\frac{\dot{\theta}}{2\pi} & (B^x_{2,1}-i B^y_{2,1})/2 & (B^x_{1,1}-i B^y_{1,1})/2 & 0  \\
 (B^x_{2,-1}+i B^y_{2,-1})/2 & \Delta E_z & 0 & (B^x_{1,1}-i B^y_{1,1})/2  \\
 (B^x_{1,-1}+i B^y_{1,-1})/2 & 0 & -\Delta E_z  & (B^x_{2,1}-i B^y_{2,1})/2 \\
 0 & (B^x_{1,-1}+i B^y_{1,-1})/2 & (B^x_{2,-1}+i B^y_{2,-1})/2 & -2(E_z-\nu_D)+\frac{\dot{\theta}}{2\pi}  \\
\end{pmatrix},
\label{eq:ham_matrix_RF2}
\end{align}
  \end{widetext}
where we kept time-dependent phases to account for shifts in resonance frequency. Corrections beyond the RWA can be included by using directly Hamiltonian~\eqref{eq:ham_matrix_RF} instead or by applying the generalized RWA introduced in Ref.~\cite{zeuchExactRotatingWave2020} and discussed in section~\ref{ssec:drivenexchange}. 

Next we define the target operation as $\left.U_\text{ideal} = e^{-i \pi \int_0^t dt^\prime [ B^x_{1,1}(t^\prime)S^x_1 + 2(\Delta f(t^\prime)-\Delta E_z(t^\prime)) S^z_2]}e^{i\theta S^z_2}\right. $ with frequency detuning $\Delta f(t)=E_z(t)+\frac{\Delta E_z(t)}{2}-\nu_D$ and find $H_\text{error}$ via Eq.~\eqref{eq:errorHam}. Our target gate operation describes single-qubit Rabi oscillations and phase shifts on the non-driven qubit. The dominant effect of off-resonant driving, a phase accumulation ($S^z_2$) on the non-driven qubit known as ac-Stark shift, is included in the target operation to keep the erroneous evolution small since it can be corrected easily via virtual $z$ gates. We identify two pairs of dominant error channels described by the operators $O_1=S_1^z\mp i S_1^y$ and $O_2=S_2^x\pm i S_2^y$ with erroneous evolutions
\begin{align}
    g_1(t) &= \Delta f(t)-\frac{\dot{\theta}(t)}{2\pi}\mp i B^y_{1,1}(t),\label{eq:1Q_g1}\\
    g_2(t) &= B^x_{2,1}(t)\pm i B^y_{2,1}(t),
    \label{eq:1Q_g2}
\end{align}
desired evolutions
\begin{align}
    \dot{f}_1(t) &=\pm \frac{h}{2} B^x_{1,1}(t), \label{eq:1Q_f1}\\
    \dot{f}_2(t) &=\pm \frac{h}{2} \left(\Delta f(t^\prime)-\Delta E_z(t) -\frac{\dot{\theta}}{2\pi}\right)
    \label{eq:1Q_f2}
\end{align}
and error rates
\begin{align}
|\text{tr}(\mathcal{E}O_1)|^2 &= S\big(\widetilde{g}_1[\nu_1 t_g]\big),\label{eq:errorHy}\\
|\text{tr}(\mathcal{E}O_2)|^2 &= S\big(\widetilde{g}_2[\nu_2 t_g]\big).
\label{eq:errorTr}
\end{align}
The first error rate describes a shift in the rotation axis ($x$-direction) of qubit 1 giving rise to $S^y_1$ and $S^z_1$ errors. The best mitigation strategy is dynamic pulse shaping via a time-dependent phase $\theta(t)$, e.g. through chirping~\cite{vandersypenNMRTechniquesQuantum2005,ShafieiResolvingSpinOrbit2013}. The second error rate describes a spin-flip of the second qubit due to off-resonant driving giving rise to $S^x_2$ and $S^y_2$ errors. The mitigation of the spin-flip errors requires either synchronization or pulse shaping. 

The conditions for the synchronization of a $R_{x,y}(\pi)$ gate on qubit $j$ affecting qubit $i$ ($i\neq j$) with a rectangular pulse shape is (see also Ref.~\cite{heinzCrosstalkAnalysisSinglequbit2021}) 
\begin{align}
     B^x_{i,1} = \frac{2 n+1}{2m}\sqrt{\frac{(B^x_{j,1})^2}{4}+\Delta E_z^2}
     \label{eq:synchronization1Q}
\end{align}
with integer $m$ and $n$. The condition for a synchronized $R_{x,y}(\frac{\pi}{2})$ gate is given by the substitution $n\rightarrow 2n$. For minimal gate time assuming an equally strong global drive $B^x_{i,1}=B^x_{j,1}$ the synchronization condition for a $R_{x}(\frac{\pi}{2})$ gate is simplified to
\begin{align}
    t_g = \frac{\sqrt{16 m^2-1}}{4\Delta E_z}
    \label{eq:synchronization1Qb}
\end{align}
which corresponds exactly to the minima in Fig.~\ref{fig:main_figure_1}~(b).

For static pulse shaping, $\theta(t)=0$ and $B^y_{i,j}=0$, constant drive frequency $\nu_D=E_z\pm\Delta E_z$, and negligible shift in resonance frequency $\Delta E_z(t)=\Delta E_z$ the first error rate Eq.~\eqref{eq:errorHy} vanishes and we only need to minimize Eq.~\eqref{eq:errorTr} which is simplified to
\begin{align}
    |\text{tr}(\mathcal{E}O_2)|^2 &= S\big(B^x_{2,1}[\Delta E_z t_g]\big).
    \label{eq:error_rate_1Q}
\end{align}
Optimal pulse shapes for a $R_{x}(\frac{\pi}{2})$ gate are then given by $B^x_{2,1}(t)= \frac{1}{4} w(t)$, where we use the normalized window $\int_0^{t_g} w(t) dt = 1$. Fig.~\ref{fig:main_figure_1}~(c) displays the simulated infidelity for different pulse shapes as a function of gate time $t_g$.

Under the same assumptions such as constant drive frequency $\nu_D=E_z\pm\Delta E_z$ and negligible shift in resonance frequency $\Delta E_z(t)=\Delta E_z$, dynamic pulse shaping provides even faster gate times with small errors (see Fig.~\ref{fig:main_figure_1})~(c)-(d). Applying the DRAG method~\cite{motzoiSimplePulsesElimination2009} Eqs.~\eqref{eq:errorHy}-\eqref{eq:errorTr} combined with a cosine shape, the optimized dynamic pulse shape is (see Fig.~\ref{fig:main_figure_1}~(d))
\begin{align}
    B^x_{2,1}(t) &= \frac{1}{4} w(t) \left(1-\frac{5}{5+(4\Delta E_z t_g)^2}\right),\label{eq:1QdragX}\\
    B^y_{2,1}(t) &= -\frac{\dot{B}^x_{2,1}(t)}{\Delta E_z},\label{eq:1QdragY}\\
    \dot{\theta}(t) &= - \frac{\left(\dot{B}^x_{2,1}(t)\right)^2}{\Delta E_z}\label{eq:1QdragZ}.
\end{align}
The renormalization of the drive amplitude is due to the additional power in driving. The optimized pulse shape for driving qubit 2 is given by substituting $\Delta E_z\rightarrow -\Delta E_z$ in Eqs.~\eqref{eq:1QdragX}-\eqref{eq:1QdragZ}.

\subsection{Exchange-based two-qubit \textsc{cz} gate}
A crucial condition for high-fidelity two-qubit \textsc{cz} gates is an adiabatic turn on/off or pulse of the exchange interaction, which can give rise to substantial errors if violated~\cite{xueQuantumLogicSpin2022a}. While in principle an echo pulse sequence allows to suppress non-adiabatic errors~\cite{russHighfidelityQuantumGates2018} for a \textsc{cz} gate, the echo pulse is often inconvenient and introduces additional noise through the longer gate times. 

Starting from Hamiltonian~\eqref{eq:ham_matrix_RF} we quickly notice that the exchange interaction only affects the odd parity states $\lbrace\ket{\uparrow\downarrow},\ket{\downarrow\uparrow}\rbrace$ (see appendix~\ref{method:exchangevoltage}). Without loss of generality, this allows us to project the full dynamics on a two-level system taking into account the global phase of this subspace. Introducing a new set of Pauli operators
$\sigma_x=\ket{\uparrow\downarrow}\bra{\downarrow\uparrow}+\ket{\downarrow\uparrow}\bra{\uparrow\downarrow}$, $\sigma_y=-i \ket{\uparrow\downarrow}\bra{\downarrow\uparrow}+i\ket{\downarrow\uparrow}\bra{\uparrow\downarrow}$, and $\sigma_z=\ket{\uparrow\downarrow}\bra{\uparrow\downarrow}-\ket{\downarrow\uparrow}\bra{\downarrow\uparrow}$, we find 
\begin{align}
H_{\text{sub}}=\frac{h}{2}\big( - J + \Delta E_z \,\sigma_z+J\, \sigma_x\big).
\label{eq:HheisSub}
\end{align}
We define the ideal target operation as $\left. U_\text{ideal}= e^{i\pi\int_0^t J(t^\prime) dt^\prime} e^{-i \pi\int_{0}^{t}dt^\prime \nu_\text{j}(t^\prime)\widetilde{\sigma}_z} \right.$, where $\widetilde{\sigma}_{x,z}=U_\text{ST}^\dagger\sigma_{x,z}U_\text{ST}$ 
with eigenenergies $\nu_\text{j}(t)=\sqrt{\Delta E_z(t)^2+J(t)^2}$ and unitary $U_\text{ST}=e^{-\frac{i}{2} \tan^{-1}\left(\frac{J(t)}{\Delta E_z (t)}\right)\sigma_y}$ diagonalizes Hamiltonian~\eqref{eq:HheisSub}. Therefore, our target operations describes the adiabatic phase evolution due to the exchange interaction. We identify a single dominant error channel described by the spin-flip operator $O=\ket{\widetilde{\uparrow\downarrow}}\bra{\widetilde{\downarrow\uparrow}}$, where $\ket{\widetilde{\uparrow\downarrow}}$ and $\ket{\widetilde{\downarrow\uparrow}}$ are the eigenstates of Hamilitonian~\eqref{eq:HheisSub}. The erroneous and targeted time evolutions are
\begin{align}
g(t) &= -h\frac{\Delta E_z\, \dot{J}-\dot{\Delta E_z} J}{4\pi\nu_\text{j}^2}, \\
\dot{f}(t) &= \frac{h}{2} \nu_\text{j}.
\end{align} 
For small exchange  $J(t)\ll \Delta E_z$ and constant Zeeman splitting $\Delta E_z(t)\approx \Delta E_z$ we can simplify $g(t)\propto \dot{J} (t)$ and $\nu_j\approx \Delta E_z$ to find the error rate
\begin{align}
P_{\ket{\uparrow\downarrow}\rightarrow\ket{\downarrow\uparrow}} \sim S\big(\dot{J}[\Delta E_z t_g]\big).
\label{eq:error_rate_2Q}
\end{align}
Remarkably, this optimization condition for an adiabatic \textsc{cz} gate is identical to the condition for minimizing single-qubit crosstalk in Eq.~\eqref{eq:error_rate_1Q} under the replacement $B^x_{2,1}\rightarrow\dot{J}$ with the same invariant $t_g\Delta E_z$.
The conditions for the synchronization of a \textsc{cz} gate with a rectangular pulse shape and minimal time $t_g$ is~\cite{burkardPhysicalOptimizationQuantum1999}
\begin{align}
    t_g = \frac{\sqrt{4 m^2-1}}{2\Delta E_z(v_B)}
\end{align}
with integer $m$. Here $\Delta E_z(v_B)$ is the difference in resonance frequency during the pulse. Note that this is equivalent to the synchronization condition of a $R_{x,y}(\pi)$ gate (see Eq.~\eqref{eq:synchronization1Qb}).

While dynamic pulse shaping is not possible for a dc pulse (no two independent control axes which are orthogonal to each other) static pulse shaping is sufficient to get low error rates.
Due to the non-linear relation between barrier voltage and exchange interaction (see Appendix~\ref{method:exchangevoltage}) the optimal pulse shape for the barrier voltage pulse shape $v_B(t)$ is then given by 
\begin{align}
    v_B(t)&=\frac{1}{\alpha}\text{log}\left(\frac{\sqrt{J(t)/J_\text{sat}}}{|1-J(t)/J_\text{sat}|}\right)\label{eq:exchange_jnv_1},\\
    J(s(t))&=\tan\left[2 A w(t_g s)+\tan^{-1}\left(\frac{J(0)}{\Delta E_z(0)}\right)\right]\Delta E_z\label{eq:exchange_jnv_3},\\
\end{align}
using the relation between real-time $t$ and time $s$. The amplitude $A$ is given by the conditional phase condition $\int_0^{t_g} J(t^\prime) dt^\prime = 1/2$.
For  $J(t)\ll \Delta E_z$ and constant Zeeman splitting $\Delta E_z(t)\approx \Delta E_z$ we find
\begin{align}
    v_B(t)&=\frac{1}{2 \alpha}\text{log}\left(\frac{w(t)}{2 J_0 t_g}+1\right)\label{eq:exchange_jnv_2},
\end{align}
where we use $t=t_g s$ and the normalized window $\int_0^{t_g} w(t) dt = t_g$. 

\subsection{Exchange-based two-qubit resonant \textsc{swap} gate}
\label{ssec:drivenexchange}
Another set of two-qubit gates can be accessed by driving the exchange interaction directly at the $\ket{\widetilde{\uparrow\downarrow}}\Longleftrightarrow\ket{\widetilde{\downarrow,\uparrow}}$ resonance frequency $\nu_\text{ST}=\sqrt{\Delta E_z^2+J_0^2}$~\cite{sigillitoCoherentTransferQuantum2019,vanriggelenPhaseFlipCode2022}. For the resonant \textsc{swap} gate, errors from violations of the rotating wave approximation due to $t_g\sim\unit[100]{ns}$ compared to $\nu_\text{ST}\sim\unit[100]{MHz}$ and the influence of higher harmonics due to the non-linear voltage-exchange relation are no longer negligible. In general driving the barrier voltage $v_B(t)=v_{B,0}+v_{B,1}(t)\cos(2\pi \nu_\text{ST}t+\theta_j)$ gives rise to
\begin{align}
J(v_B) =& \mathcal{J}_0(v_B)+ \sum_{k> 0} 2\mathcal{J}_{k}(v_B) \cos(2\pi \nu_\text{ST} k t+ k \theta)  ,\label{eq:drivenExchange}\\
\Delta E_z(v_B) =& \Delta \mathcal{E}_{z,0}(v_B) \nonumber\\ 
&+\sum_{k> 0} 2\Delta \mathcal{E}_{z,k}(v_B)  \cos(2\pi \nu_\text{ST} k t+ k \theta) ,\label{eq:drivenZeeman}
\end{align}
where we have expressed $J$ and $\Delta E_z$ in terms of Fourier coefficients with respect to the drive frequency $\nu_\text{ST}$. Since tuning the barrier voltage typically accompanies a change in resonance frequency~\cite{xueQuantumLogicSpin2022a}, we consider this in our model $\Delta E_z \rightarrow \Delta E_z(v_{B}(t))$ with $\Delta E_z(v_{B,0})=\Delta E_z$. 

Without loss of generality, the dynamics is again projected on the odd-parity subspace spanned by $\lbrace\ket{\uparrow\downarrow},\ket{\downarrow\uparrow}\rbrace$ and described by Hamiltonian~\eqref{eq:HheisSub}. In order to simplify the Hamiltonian we perform a double basis transformation $U=e^{-i (\pi\nu_\text{ST}t+\theta_j/2) \widetilde{\sigma}_z}e^{-\frac{i}{2} \tan^{-1}\left(\frac{J_0}{\Delta E_z}\right)\sigma_y}$ before we apply our framework. The first transformation diagonalizes $H_\text{sub}(t=0)$ and the second moves us into the rotating frame with respect to the driving frequency and driving phase. The transformed and rotated Hamiltonian reads
\begin{align}
    H_\text{ac} =& -\frac{h }{2}J(t) \nnb
    &+\frac{h}{2} \left(
    \frac{J_0\, J(t)+\Delta E_z\, \Delta E_z(t)}{\nu_\text{ST}}-\nu_\text{ST}-\frac{\dot{\theta}_j}{2\pi}
\right)\widetilde{\sigma}_z \nonumber\\
&+ \frac{h}{2}\left(\frac{\Delta E_z J(t)-J_0 \Delta E_z(t) }{\nu_\text{ST}}e^{2\pi i \nu_\text{ST} t +i \theta_j}\right) \widetilde{\sigma}_+ + h.c.
    \label{eq:acExchangeHam}
\end{align}
With this transformation, the target operation is simply expressed as $\left.U_\text{ideal} = e^{-i \pi \int_0^{t}dt^\prime \cos(2\pi \nu_\text{ST} t + \theta_j) (\Delta E_z J(t)-J_0 \Delta E_z(t))\widetilde{\sigma}_x/\nu_\text{j}(t) }\right.$ which describes a flip between the eigenstates, i.e. $\ket{0}\sim\ket{\uparrow\downarrow}$ and  $\ket{1}\sim\ket{\downarrow\uparrow}$ for $\Delta E_z\gg J_0$. We identify a single pair of error rates described by the operators $O=\widetilde{\sigma}_z\pm i \widetilde{\sigma}_y$. The erroneous and targeted time evolution are
\begin{align}
g(t) &= \frac{J_0\,J(t)+\Delta E_z\, \Delta E_z(t)}{\nu_\text{ST}}-\nu_\text{ST}-\frac{\dot{\theta}_j}{2\pi}  \nonumber\\
&\mp i \sin(2\pi \nu_\text{ST} t + \theta_j) \frac{\Delta E_z J(t)-J_0 \Delta E_z(t)}{2\nu_\text{ST}},\label{eq:2Q_ac_g_0} \\
\dot{f}(t) &= h\cos(2\pi \nu_\text{ST} t + \theta_j) \frac{\Delta E_z J(t)-J_0 \Delta E_z(t)}{2\nu_\text{ST}}.\label{eq:2Q_ac_f_0}
\end{align}

\paragraph{General case}
In the general case no closed-form analytical expressions can be derived for Eq.~\eqref{eq:drivenExchange} and one has to rely on numerical techniques. Additionally, static and dynamic pulse shaping is also not directly possible due to $g(t)$ being complex but can be made possible by focusing on the error rate. We start by separating $g(t)=g_R(t)+i g_I(t)$ where $g_R$ and $g_I$ are real functions and then perform integration by parts
\begin{align}
    |\text{tr}(\mathcal{E}\,O)|^2 &= \left|\int_0^{t_g}[g_R(t)+i g_I(t)]e^{i f(t)/\hbar}dt  \right|^2\\
    &= \left|\int_0^{t_g}\Big[g_R(t)+ \frac{\dot{f}(t)}{\hbar} \int_0^{t}g_I(t^\prime)dt^\prime\Big] e^{i f(t)/\hbar}dt  \right|^2.
    \label{eq:optimized_shape_ac}
\end{align}
Here, we restrict ourselves to solutions where the limits of integration vanish at $t=0,t_g$. Unfortunately, we cannot apply the next (optional) step of our framework and use the substitution~\eqref{eq:substitution} since $\dot{f}=0$ gives rise to non-unique solutions. We note that our full framework can still be applied if we apply a (generalized) RWA~\cite{zeuchExactRotatingWave2020} directly on Hamiltonian~\eqref{eq:acExchangeHam} to remove the oscillating components. Dynamic pulse shaping is however possible as we directly arrive at an integro-differential equation for the phase $\theta_j$ by plugging in Eqs.~\eqref{eq:2Q_ac_g_0}~and~\eqref{eq:2Q_ac_f_0} whose solution provides an optimized $\theta_j(t)$.

\begin{figure*}
	\begin{center}
		\includegraphics[width=1.\textwidth]{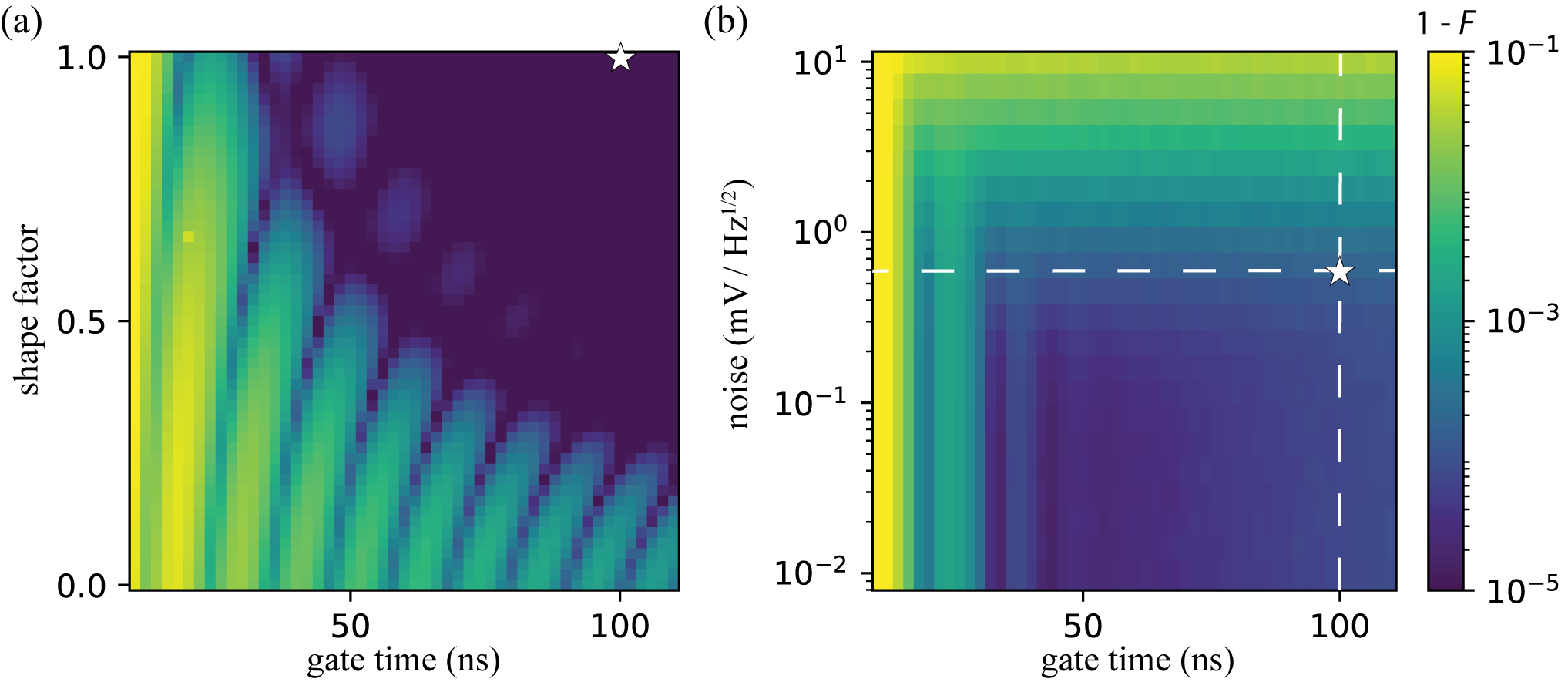}
	\end{center}
	\caption{(a) Simulated noiseless average gate infidelity $1-F$ for a \textsc{cz} gate as a function of pulse length $t_g$ and the shaping factor $\lambda$ of the Tukey window Eq.~\eqref{eq:Tukey_window}. The star highlights the pulse shape used in Ref.~\cite{xueQuantumLogicSpin2022a} demonstrating $F=\unit[99.6]{\%}$ in isotopically enriched silicon quantum dots. (b) Simulated infidelity $1-F$ for a \textsc{cz} gate as a function of pulse length $t_g$ and charge noise amplitude $A$ of the virtual voltage using a cosine shape ($\lambda = 1$). The dashed lines highlight the value of extracted charge noise and pulse duration from Ref.~\cite{xueQuantumLogicSpin2022a}.}
	\label{fig:main_figure_2}
\end{figure*}
\paragraph{Exponential exchange}
We will now consider the case of an exponential interaction $J(v_B)\propto e^{2\alpha v_B} $ and a linear frequency shifts, $\Delta E_z(v_B) = \Delta E_z + \beta\,  (v_{B}-v_{B,0})$. This regime is experimentally accessed in Refs.~\cite{sigillitoCoherentTransferQuantum2019,vanriggelenPhaseFlipCode2022}. In this limit, exact analytic expressions can be derived since the Fourier coefficients are given by
\begin{align}
    \mathcal{J}_{k}(v_{B})=J_0\mathcal{I}_k(2\alpha (v_{B}-v_{B,0}))
\end{align}
for $k=0,\pm 1$, where $\mathcal{I}_k$ denotes the modified Bessel function of order $k$. 
In this special case the optimized dynamic pulse shape can then be simplified to
\begin{align}
    v_{B,1}(t)&=\frac{1}{2 \alpha}\text{log}\left(\frac{w(t)}{4 J_0 t_g}+1\right) \cos(2\pi \nu_\text{ST}+\theta_j(t))\label{eq:exchange_ac}
\end{align}
with the dynamic phase
\begin{align}
    \theta_{j}(t) =&\phantom{+} 2\pi\int_0^t dt^\prime \frac{J_0\,\mathcal{J}_0(v_{B}(t^\prime))+\Delta E_z^2-\nu^2_\text{ST}}{\nu_\text{ST}}\nonumber\\
    &+ 2\pi\int_0^t dt^\prime \frac{\Delta E_z J_1(t^\prime)- J_0 \beta\,v_{B,1}(t^\prime) }{\nu_\text{ST}^2}\nonumber\\
    &\phantom{+2\pi\int_0^t}\times \bigg[\frac{\Delta E_z J_0(t^\prime)}{2\pi\nu\st{ST}^3}+\frac{\Delta E_z J_1(t^\prime)}{4\pi\nu\st{ST}^3}\nonumber\\
    &\phantom{+2\pi\int_0^t\times \bigg[}-\frac{J_0 \beta\,v_{B,1}(t^\prime)}{8\pi\nu\st{ST}^3}-\sum_{k=2}^\infty\frac{\Delta E_z J_k(t^\prime)}{(k^2-1)\pi\nu\st{ST}^3}\bigg].
    \label{eq:phase_correction}
\end{align}
%\textbf{Check formula on correctness. Last line should include $\mathcal{J}_0$? Second line should be combined with third line?}
The first term originates from the conventional RWA and compensates the shift in resonance frequency $h\sqrt{\Delta E_z^2+J^2}$ due to non-linear exchange. The remaining term describes the driving-induced shift of the rotation angle and can be derived with the help of a generalized RWA~\cite{zeuchExactRotatingWave2020} (see Appendix~\ref{method:generalRWA}).

\section{Performance}
\label{sec:results}
In this section we show that the aforementioned techniques lead to high-fidelity single- and two-qubit gates. We benchmark the gates by computing the time-evolution of an input state $\Psi(t)$ by step-wise integration of the Schroedinger equation
\begin{align}
    i\hbar \dot{\Psi}(t) = H(t) \Psi(t),
\end{align}
where $H(t)$ is the exact Hamiltonian~\eqref{eq:ham_matrix_RF} in the rotating frame without neglecting the counter-rotating terms.
Noise is added in two ways into the dynamics. Because of the slow dynamics compared to the gate times $<\unit[1]{\mu s}$, magnetic noise affecting the single spins is simulated by a quasi-static shift of the qubit resonance frequencies $B_{z,i}$. Charge noise is simulated using colored noise with a spectral density $S(f) = \frac{A^2}{2\pi f}$ using the Fourier Filter method~\cite{yangAchievingHighfidelitySinglequbit2019,koskiStrongPhotonCoupling2020}. To simulate the performance at the symmetric operation point~\cite{martinsNoiseSuppressionUsing2016,reedReducedSensitivityCharge2016}, where decoherence is greatly improved, voltage fluctuations predominantly affect the exchange interaction non-linearly via the barrier potential $v_B(t)=v_{B,0}(t) + \delta v_B(t)$.

\subsection{Resonant single-qubit gates}
 The most common gates in a multi-qubit algorithm are single-qubit gates. Therefore, each large-scale qubit device needs consistently small single-qubit error rates to fulfill the requirements for error correction~\cite{raussendorfFaultTolerantQuantumComputation2007,lidarQuantumErrorCorrection2013}. A frequently reported number are infidelities of $1-F<0.5-1\%$ for the physical gate operations~\cite{raussendorfFaultTolerantQuantumComputation2007,fowlerSurfaceCodesPractical2012}, although based on many assumption such as good initialization and readout and uncorrelated errors. Our simulations show that sufficiently small infidelities are within reach on state-of-the-art qubit devices using our framework. The simplest technique, synchronization of the Rabi-frequencies~\cite{russHighfidelityQuantumGates2018,heinzCrosstalkAnalysisSinglequbit2021} already shows extremely small infidelities in our simulations. However, low-pass filters in the signal transfer to limit high-frequency noise are a detrimental error source for the synchronization technique as seen in Fig.~\ref{fig:main_figure_1}-(b) that shows the infidelity with (blue) and without (orange) a \unit[150]{MHz}-Butterworth filter. The filter significantly reduces the dips as well as shifts the minima. 
On the other hand our simulations (see Fig.~\ref{fig:main_figure_1}-(c)-(d)) show that high fidelity operations can still be reached using static or dynamic pulse shaping to reach infidelities as small as $10^{-4}$ for gate times in the order of $t_g=\unit[25]{ns}$ with a frequency separation of $|f_{Q,1}-f_{Q,2}|=\unit[100]{MHz}$. Note, that this on the one hand directly implies that gate times $t_g=\unit[250]{ns}$ are required if the frequency separation is reduced to $|f_{Q,1}-f_{Q,2}|=\unit[10]{MHz}$ as proposed in some architectures, e.g. Ref.~\cite{vandersypenInterfacingSpinQubits2017}, but on the other hand a larger frequency separation allows for faster high-fidelity gate operations~\cite{lawrieSimultaneousDrivingSemiconductor2021a,millsTwoqubitSiliconQuantum2022} due to the infidelity being invariant for $t_g\times |f_{Q,1}-f_{Q,2}|$. In the former case dynamic pulse shaping, such as the derivative removal by adiabatic gate (DRAG) protocol~\cite{motzoiSimplePulsesElimination2009,luceroReducedPhaseError2010} allows for an additional improvement.

\begin{figure*}
	\begin{center}
		\includegraphics[width=1.\textwidth]{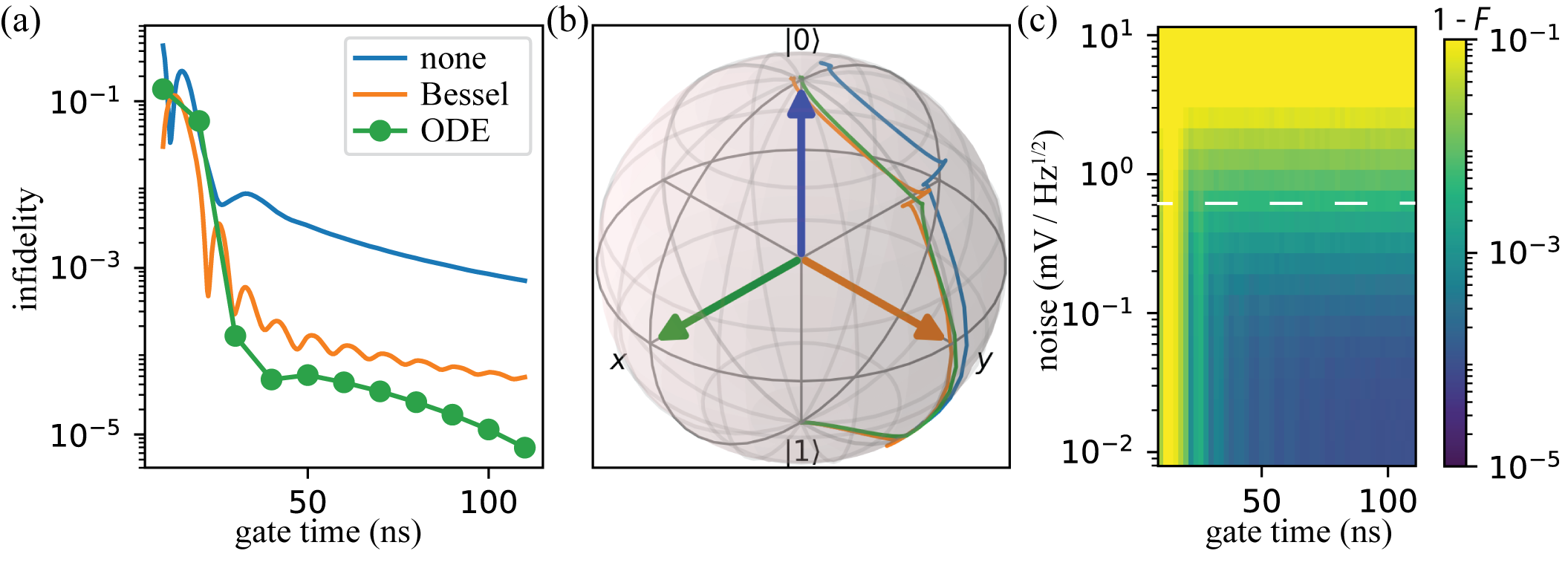}
	\end{center}
	\caption{(a) Simulated average gate infidelity $1-F$ of a \textsc{swap}-class gate as a function of pulse length $t_g$ using no dynamic pulse shaping (blue), dynamic pulse shaping based on Bessel functions Eq.~\eqref{eq:phase_correction} (orange), and dynamic pulse shaping based on a ODE approximation of Eq.~\eqref{eq:optimized_shape_ac}  (green). The simulation using ODE was performed with Mathematica. (b) Trajectories of the optimized pulse shapes for a $t_g=\unit[40]{ns}$ projected on the Bloch sphere. Only dynamic pulse shaping based on the ODE approximation gives rise to a high fidelity state flip. (c) Simulated gate infidelity $1-F$ as a function of pulse length $t_g$ and charge noise amplitude $A$ of the virtual voltage using a cosine shape and dynamic pulse shaping based on Bessel functions. The dashed line highlights the value of charge noise of the virtual voltage extracted in Ref.~\cite{xueQuantumLogicSpin2022a}.}
	\label{fig:main_figure_3}
\end{figure*}

\subsection{Exchange-based two-qubit \textsc{cz} gate}
Many experiments on single-spin qubits use the universal \textsc{cz} gate~\cite{burkardPhysicalOptimizationQuantum1999,meunierEfficientControlledphaseGate2011}
as their native high-fidelity two-qubit gate~\cite{veldhorstTwoqubitLogicGate2015,watsonProgrammableTwoqubitQuantum2018,xueBenchmarkingGateFidelities2019,xueQuantumLogicSpin2022a,madzikPrecisionTomographyThreequbit2022a,millsTwoqubitSiliconQuantum2022} due to its simplicity and potential for scaling to larger arrays~\cite{hendrickxFourqubitGermaniumQuantum2021,philipsUniversalControlSixqubit2022}. The \textsc{cz} gate 
\begin{align}
    U_\text{\textsc{cz}} &=e^{-i \Phi_\textsc{cz} \left(S_{1,z}S_{2,z}-\frac{1}{4}\right) }\\
    &\equiv \text{diag}(1,1,1,e^{i \Phi_\textsc{cz}})
\end{align}
with $\Phi_\textsc{cz} = 2\pi\int_0^{t_g}J(t)dt$ can be directly acquired using only single-qubit phase gates~\footnote{Can be implemented virtually using only software phase shifts} from the adiabatic phase evolution under the exchange interaction and can be transformed into a \textsc{cnot} gate using two single-qubit $R_{y}(\frac{\pi}{2})$ gates. For the \textsc{cz} gate the exchange interaction $J(t)$ is pulsed picking up a conditional phase $\Phi_\textsc{cz}=  (2n+1)\pi $ with integer $n$. Since the \textsc{cz} gate is intended to be adiabatic by nature, it is also directly (linearly) susceptible to low-frequency noise acting on the resonance frequencies. However, we claim that a substantial error in previous realizations~\cite{veldhorstTwoqubitLogicGate2015,petitDesignIntegrationSinglequbit2022} is due to violations of the adiabaticity condition which is more severely impacted due to the non-linear voltage-exchange relation and gives rise to bit-flip errors. Here, the adiabaticity condition is with respect to the frequency difference $|f_{Q,2}-f_{Q,1}|$ between the two spin qubits. Since the coherent errors of a \textsc{cz} two-qubit and resonant single-qubit gates are related and $t_g\times |f_{Q,1}-f_{Q,2}|$ is an invariant in the simulations, we know that the \textsc{cz} gate can, for example, be further improved using the Kaiser window or increasing the separation in qubit frequency. Fig.~\ref{fig:main_figure_2}~(a) shows the simulated infidelity of a \textsc{cz} gate as a function of gate time $t_g$ and the proportionality factor $\alpha\in [0,1]$ from the Tukey window, see Eq.~\eqref{eq:Tukey_window}. The oscillating infidelity as a function of exchange corresponds to an interference pattern of the diabatic contributions identical to the one observed in Fig.~\ref{fig:main_figure_1}~(b)~and~(c). On the other hand a small frequency separation severely limits the performance since long gate times $t_g=\unit[300]{ns}$ are required if the frequency separation is reduced to $|f_{Q,1}-f_{Q,2}|=\unit[10]{MHz}$~\cite{petitUniversalQuantumLogic2020,yangOperationSiliconQuantum2020a}. In situations with a small frequency separation splitting a non-adiabatic \textsc{cz} gate can be realized as shown theoretically~\cite{burkardPhysicalOptimizationQuantum1999} and demonstrated experimentally~\cite{petitDesignIntegrationSinglequbit2022} using the synchronization condition. Our simulations confirm such high-fidelity gates in Fig.~\ref{fig:main_figure_2}~(a) for $\lambda=0$, where the Tukey pulse corresponds to a rectangular pulse. However, the diabatic implementation requires a precise timing, is sensitive to pulse imperfections such as filter effects, and is prone to dephasing due to low-frequency noise. 

We now compare our results to recent experiments demonstrating high-fidelity \textsc{cz}-gate operations with infidelities $1-F=4\times 10^{-3}$~\cite{xueQuantumLogicSpin2022a} and $1-F=2\times 10^{-3}$~\cite{millsTwoqubitSiliconQuantum2022}. Both experiments are performed in isotopically enriched silicon quantum dots which are prone to (low-frequency) charge noise. Low-frequency noise couples to the \textsc{cz}-gate through the diagonal matrix elements $S^z_{1,(2)}$ and $S^z_{1}S^z_{2}$ via the qubit frequencies $\Delta E_z$ and via the exchange interaction $J$. In Fig.~\ref{fig:main_figure_2} we highlighted the pulse shape, expected noise level, and pulse duration (star) of the extracted parameters in Ref.~\cite{xueQuantumLogicSpin2022a}. Our simulation predicts coherent errors as low as $1-F=10^{-5}$ and total errors of $1-F=2\times 10^{-4}$ are achievable. We speculate that the discrepancy of simulations and experiment is related to heating effects~\cite{undsethNonlinearResponseCrosstalk2022}. While we haven't performed simulations using the parameters in Ref.~\cite{millsTwoqubitSiliconQuantum2022} we can nevertheless predict the coherent infidelity using the invariant $t_g\times |f_{Q,1}-f_{Q,2}|$. The conversion results in a rectangular pulse with an effective gate duration $\widetilde{t}_g=\unit[158]{ns}$ due to the frequency difference $|f_{Q,2}-f_{Q,1}|=\unit[396]{MHz}$ (derivation see appendix~\ref{appendix:mills_coherent} for details). Our simulations predict a coherent infidelity of $1-F=5\times 10^{-4}$ which is increased to $1-F=8\times 10^{-4}$ if incoherent noise sources are included.

Fig.~\ref{fig:main_figure_2}~(b) shows the \textsc{cz} gate infidelity in general as a function of charge noise amplitude $A$ and gate time $t_g$ showing the importance of the interplay between coherent and incoherent errors. Too fast gates suffer from coherent errors while too slow gates are prone to incoherent errors with an optimum depending on the charge noise amplitude. 
Taking into account realistic values~\cite{xueQuantumLogicSpin2022a} (see dashed line in Fig.~\ref{fig:main_figure_2})~(b)) one can clearly observe that coherent errors from non-adiabicity are dominating for $t_g<\unit[40]{ns}$ and infidelities as low as $1-F=10^{-4}$ are possible. 

\subsection{Exchange-based two-qubit resonant \textsc{swap} gates}

While arbitrary single-qubit gates combined with \textsc{cz} form a universal gate set for quantum circuits, it is often more efficient to include additional gates into the gate set. A frequently neccessary gate in qubit architectures with next-neighbor couplings only is the \textsc{SWAP} gate, as it enables long-range qubit-qubit communication as well as read-out~\cite{vandersypenInterfacingSpinQubits2017,sigillitoCoherentTransferQuantum2019,philipsUniversalControlSixqubit2022,vanriggelenPhaseFlipCode2022}. The \textsc{swap}-class gate
\begin{align}
    U_\text{\textsc{swap}}(\phi)=\begin{pmatrix}
      1 & 0 & 0 & 0 \\
      0 & 0 & e^{i\Phi_\textsc{swap}} & 0 \\
      0 & e^{i\Phi_\textsc{swap}} & 0 & 0 \\
      0 & 0 & 0 & 1 \\
    \end{pmatrix}.
    \label{eq:swapclass}
\end{align}
can be either accessed directly through a diabatic exchange pulse for small frequency differences $|f_{Q,2}-f_{Q,1}|\ll J$~\cite{meunierEfficientControlledphaseGate2011,petitDesignIntegrationSinglequbit2022} or in general by driving the exchange interaction at the frequency difference $|f_{Q,2}-f_{Q,1}|$ between the corresponding qubits. Here we defined the \textsc{swap}-class gate via the additional phase $\Phi_\textsc{swap}=\pi \int_0^{t_g} J(t)dt$. For $\Phi_\textsc{swap}=n\pi$ we recover the classical \text{swap}-gate while the i\textsc{swap}, $\Phi_\textsc{swap}=(2n+1)\pi/2$, maximally entangles the qubits during the swap. Fig.~\ref{fig:main_figure_3}~(a) shows the simulated infidelity of a \textsc{swap}-class gate
for different static and dynamic pulse shapes assuming perfect phase compensation. Due to the non-linear exchange interaction and short target gate times $|f_{Q,2}-f_{Q,1}|\sim 1/t_g$ dynamic pulse shaping greatly enhances the performance of the gates. For situations when the explicit phase $\Phi_\textsc{swap}$ matters, e.g., compiling Clifford gates using i\textsc{swap} gate, we provide two methods. The simplest method to obtain for example a i\textsc{swap} gate is to append a \textsc{cz} gate such that $\Phi_\textsc{swap}+\Phi_\textsc{cz}/2=(2n+1)\pi/2$~\cite{sigillitoCoherentTransferQuantum2019}.
Remarkably, one can also perform the compensation \textsc{cz} gate simultaneously with the \textsc{swap} gate by combining an ac and dc control signal. This has the advantage of a faster gate time at the cost of additional calibrations. 

\begin{figure}
	\begin{center}
		\includegraphics[width=1.\columnwidth]{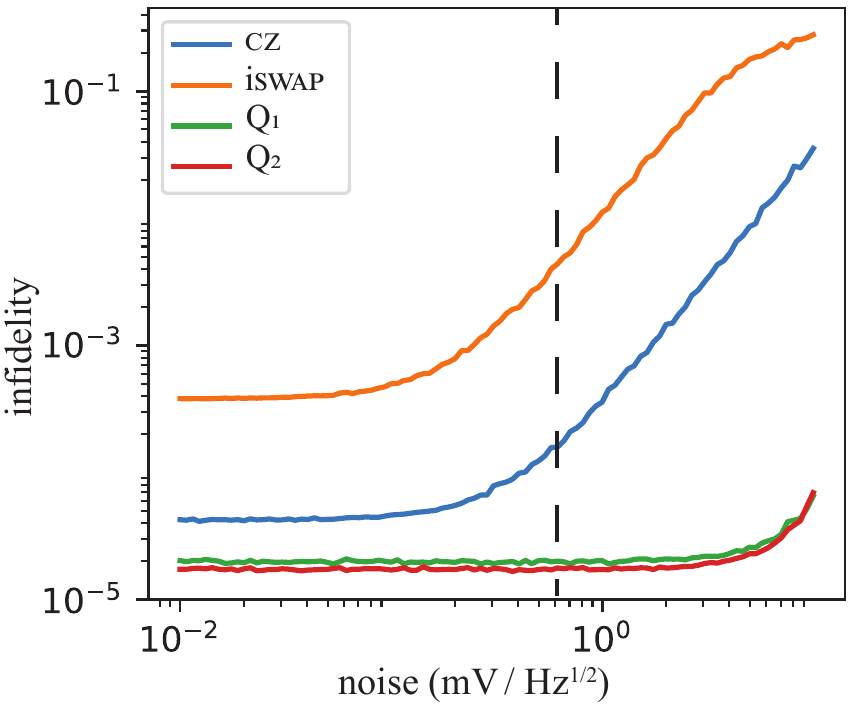}
	\end{center}
	\caption{Simulated infidelity $1-F$ of the \textsc{cz} gate, i\textsc{swap} gate, and single-qubit $R_{x,1}(\pi/2)$ gates on qubit $Q_1$ and $Q_2$ as a function of charge noise amplitude $\sigma_{v_B}$ of the virtual voltage with a constant level of nuclear noise. All gate times are $t_g=\unit[35]{ns}$. The dashed line highlights the value of charge noise of the virtual voltage extracted in Ref.~\cite{xueQuantumLogicSpin2022a}.}
	\label{fig:benchmark_all}
\end{figure}

The \textsc{swap}-class is directly (linear) susceptible to low-frequency noise coupling in via the exchange interaction $J$.
Additionally, both discussed implementations have in common that they require careful calibration to compensate for the adiabatic phase acquisition from exchange making them (at least) equally susceptible to low-frequency charge noise as the conventional \textsc{cz} gate. 
Fig.~\ref{fig:benchmark_all} compares the infidelity of the different two-qubit gate implementations discussed in this paper. It is clearly visible that the \textsc{cz} gate always outperforms the \textsc{swap}-class gate. The lower fidelity of the \textsc{swap} gate is due to the overall larger conditional phase picked up, $\Phi_\textsc{swap}\approx 2\Phi_\textsc{phase}$.

\subsection{Conclusion}
In this work we have presented a framework which allows us to characterize unitary errors and how to suppress these for various basic gate operations for spin qubits. Unitary errors mostly arise due to violations of approximations such as the rotating wave approximation, larger system sizes in the form of crosstalk, and through non-linear transfer functions of the input signal. Our numerical simulations show that for state-of-the-art experiments unitary errors can indeed be the limiting factor. 

Explicitly, we used our framework to obtain optimized pulse shapes for a resonantly driven single-qubit gates and exchange-based dc or ac gates on single-spin qubits. These techniques have been successfully implemented and enabled a two-qubit \textsc{cz} gate with fidelity $F=\unit[99.6]{}$~\cite{xueQuantumLogicSpin2022a}. We have also shown that the optimized static pulse shapes for single-qubit gates and \textsc{cz} two-qubit gates are identical and only depend on the qubit frequency separation. This possibly allows for a direct on-chip integration of the control electronics with little memory requirements. The transformation of the signal to compensate the exponential relationship between voltage and exchange interaction is possible using efficient digital algorithms or an analog logarithmic element. 

Our framework and also all of the presented optimized pulse shapes are directly applicable to different platforms. To suppress coherent errors even further, higher-order Magnus expansion terms can be taken into account~\cite{ribeiroSystematicMagnusBasedApproach2017}. In our formalism this would correspond to not only minimizing the spectral density but also minimizing higher correlations such as the bi-spectrum or multi-spectrum.

While in this work we focused on improving the performance of operations with respect to coherent errors we see no issue with expanding the formalism to also account for incoherent errors. We can think of two steps how this can be achieved. First, we can either extend our formalism to describing the time dynamics in terms of a propagator based on the Liouville superoperator instead of unitary operations~\cite{blume-kohoutTaxonomySmallMarkovian2022}. Alternatively, to account for low-frequency noise, we can combine our framework with the geometric gate formalism introduced in Ref.~\cite{barnesDynamicallyCorrectedGates2022}.

\acknowledgments
The authors greatly acknowledge the contributions of S. de Snoo to the manuscript. We thank all members of the Veldhorst and Vandersypen group, M. Mehmandoost, D. Zeuch, V. Evangelos, and Vincent Bejach for inspiring and helpful discussion.
M.R. acknowledges support from NWO under Veni grant (VI.Veni.212.223). This research was sponsored by the Army Research Office (ARO) under grant numbers W911NF-17-1-0274 and W911NF-12-1-0607. The views and conclusions contained in this document are those of the authors and should not be interpreted as representing the official policies, either expressed or implied, of the ARO or the US Government. The US Government is authorized to reproduce and distribute reprints for government purposes notwithstanding any copyright notation herein.

\section*{Data availability}
Simulation software and data analysis scripts supporting this publication are available at \doi{10.5281/zenodo.7341187}.

\appendix

% \section*{Methods}
\section{Sparse time evolution}
\label{method:sparse_evolution}
Starting from Eq.~\eqref{eq:Magnus_approx} we now derive Eq.~\eqref{eq:sparseEvolution} which is the starting point of our framework. We do this in two steps. First, the interior form of Eq.~\eqref{eq:sparseEvolution} is a direct consequence of the separation of the dynamics into desired and erroneous Hamiltonian contributions, $H=H_\text{ideal} + H_\text{rest}$, combined with the transformation into the interaction frame with respect to $H_\text{ideal}$ due to $i\hbar U_\text{ideal}^\dagger(t)\dot{U}_\text{ideal}(t)=H_\text{ideal}$. Thus we can rewrite Eq.~\eqref{eq:errorHam}
\begin{align}
    H_\text{error}&=U_\text{ideal}^\dagger(t)HU_\text{ideal}(t)-i\hbar U_\text{ideal}^\dagger(t)\dot{U}_\text{ideal}(t),\\
    &=U_\text{ideal}^\dagger(t)HU_\text{ideal}(t)-H_\text{ideal},\\
    &=U_\text{ideal}^\dagger(t)(H_\text{ideal} + H_\text{rest})U_\text{ideal}(t)-H_\text{ideal},\\
    &=U_\text{ideal}^\dagger(t)H_\text{rest}U_\text{ideal}(t)\nonumber\\
    &\phantom{=}+U_\text{ideal}^\dagger(t)H_\text{ideal}U_\text{ideal}(t)-H_\text{ideal},\\
    &=U_\text{ideal}^\dagger(t)H_\text{rest}U_\text{ideal}(t),
\end{align}
where we only inserted the two upper definitions and the commutativity between $H_\text{ideal}$ and $U_\text{ideal}$. In the next step we separate $H_\text{rest} = H_\text{rest,c}+H_\text{rest,a}$, where $H_\text{rest,c}$ is the commuting component $[H_\text{rest,c},U_\text{ideal}(t)]=0$ and $H_\text{rest,a}$ the remainder. The commuting component can be corrected independently of the ideal operation thus trivially calibrated or corrected. Without loss of generality we assume $H_\text{rest,c}=0$ otherwise we redefine $H_\text{ideal}=H_\text{ideal}+H_\text{rest,c}$ and drop the sub-indices. We can now always define a set of complex parameters $g_k (f_k)$ associated to the erroneous (ideal) dynamics to rewrite every matrix element of 
\begin{align}
    H_{\text{error},k_1,k_2}=\sum_k g_k(t) e^{i f_k(t)/\hbar}
    \label{eq:sparse_evolution_step_1}
\end{align}
as no restrictions are placed on $g_k$ and $f_k$. 
Second, the outer form of Eq.~\eqref{eq:sparseEvolution} follows directly from the lowest-order Magnus expansion. Since we assumed small errors we truncate the matrix exponential  (Eq.~\eqref{eq:Magnus_approx}) in linear order 
\begin{align}
    \mathcal{E}\approx\exp\left(-\frac{i}{\hbar}\int_0^{t_g} H_\text{error}(t^\prime) dt^\prime\right)\\
    \approx 1+\left(-\frac{i}{\hbar}\int_0^{t_g} H_\text{error}(t^\prime) dt^\prime\right)
    \label{eq:sparse_evolution_step_2}
\end{align}
Without loss of generality, we now choose the operator $O=\ket{k_1}\bra{k_2}$ since otherwise we can rewrite $O=\sum_m O_m$, restrict ourselves to a single operator component $O_m$, and sum up the components afterwards. Finally, combining Eq.~\eqref{eq:sparse_evolution_step_1} and Eq.~\eqref{eq:sparse_evolution_step_2} gives rise to the final expression Eq.~\eqref{eq:sparseEvolution} in the main text. An alternative derivation of Eq.~\eqref{eq:sparseEvolution} for a specific application case using a geometric interpretation can be found in Ref.~\cite{martinisFastAdiabaticQubit2014}.

\section{Modeling the exchange interaction}
\label{method:exchangevoltage}
Eq.~\eqref{eq:ham_matrix} of the main text is an approximation of the spin-dynamics in the low-energy subspace considering a single fermion in the left and right quantum dots, $(1,1)$ charge configuration, in the presence of small spin-orbit interaction~\cite{burkardCoupledQuantumDots1999,huHilbertspaceStructureSolidstate2000}. The origin of the spin-orbit interaction (SOI) may arise from intrinsic properties~\cite{winklerSpinorbitCouplingEffects2003} or artificial created through the deployment of micromagnets~\cite{tokuraCoherentSingleElectron2006}. Without (with negligible) SOI the low-energy dynamics of the spin can be derived starting from a Hubbard model with spin-conserving tunneling elements using a Schrieffer-Wolff approximation. Due to the Pauli exclusion principle, the spin state of a doubly occupied orbital state is always a spin singlet. Therefore, in the $(1,1)$ configuration only the singlet state, $\ket{S}=\frac{1}{\sqrt{2}}(\ket{\uparrow\downarrow}-\ket{\downarrow\uparrow})$, can hybridize and be lowered in energy. Consequently, the exchange interaction can be written as
\begin{align}
    H_\text{exchange}&=-hJ\ket{S}\bra{S}\\
    &=hJ\,(\boldsymbol{S}_1\cdot \boldsymbol{S}_2-\frac{1}{4}).
\end{align}
The dynamics of the (isotropic) exchange interaction is thus limited to the singlet-state. In the presence of a difference in qubit resonance frequencies $\Delta E_z$ the $\ket{\uparrow\downarrow}$ and $\ket{\downarrow\uparrow}$ states are energetically separated, thus coupling the singlet with the triplet $\ket{T_0}=\frac{1}{\sqrt{2}}(\ket{\uparrow\downarrow}+\ket{\downarrow\uparrow})$ state. 

The amplitude of the exchange interaction $J$ is a non-linear function of an applied (virtual) barrier voltage $v_B$. In most experiments the exchange interaction can be modelled as an exponential function~\cite{cerfontaineHighfidelityGateSet2020,vandiepenAutomatedTuningInterdot2018,panResonantExchangeOperation2020}
\begin{align}
J(v_B) &= J_\text{sat} e^{2\alpha (v_B-v_\text{off})}\\
\equiv J_0 e^{2\alpha v_B}.
\label{eq:J_sim_expression}
\end{align}
Other experiments~\cite{dialChargeNoiseSpectroscopy2013,reedReducedSensitivityCharge2016}, further indicate a saturation for large exchange values, thus, the upper expression~\eqref{eq:J_sim_expression} can be seen as an approximation for $J\ll J_\text{sat}$. A more general expression considering saturation reads~\cite{reedReducedSensitivityCharge2016}
\begin{align}
J(v_B) &= J_\text{sat}\left(\sqrt{1+e^{-2 \alpha (v_B-v_\text{off})}}-e^{-\alpha (v_B-v_\text{off})})\right)^2.
\label{eq:J_gen_expression}
\end{align}
Here $\alpha$ is the leverarm, $v_\text{off}$ is an offset which is set by the residual exchange interaction $J_\text{res}=J(0)$, and $J_\text{sat}$ describes the saturation value of the exchange interaction when the two electrons are strongly hybridized. In practice $J_\text{sat}$ can be motivated to be the singlet-triplet splitting or exchange splitting for a merged double quantum dot. 

The presence of a valley degree of freedom~\cite{rohlingUniversalQuantumComputing2012,davidEffectiveTheoryMonolayer2018} affects the exchange interaction $J$ as well as the frequency difference $\Delta E_z$. In lowest-order perturbation theory we find
\begin{align}
    \Delta \widetilde{E}_z &= \Delta E_z \left[1-\left( \frac{t^2}{U^2} + \frac{2t^2}{(E_{V,1}+E_{V,2} + 2U)^2} \right) \right]\label{eq:Correction_ValleySplitting},\\
    \widetilde{J} &= J\,\,\frac{1+\cos(\phi_{V,1}-\phi_{V,2})}{2}\label{eq:Correction_ValleyPhase},
\end{align}
where $E_{V,i}$ and $\phi_{V,i}$ are the respective valley splitting and valley phase of dot $i$.  Plugging in realistic parameters $t=\unit[20]{\mu eV}$, $U=\unit[3]{meV}$, $\Delta E_z=\unit[100]{MHz}$, and $E_V=\unit[200]{\mu eV}$ we find $\Delta E_z-\Delta \widetilde{ E}_z\approx\unit[8]{kHz}$. On contrast a finite valley phase difference significantly suppresses the exchange interaction~\cite{tariqImpactValleyOrbit2022}. For our theoretical simulations the valley phase becomes only problematic once it changes during the pulse, e.g. due to the deformation of the wave-functions as a function of applied voltages. Since the dc and ac pulsed two-qubit gates are adiabatic with respect to the qubit frequencies $f_{Q,1,Q,2}=E_z\pm\Delta E_z/2$
and in most devices also with respect to the valley splitting $E_V$ the gates are only slightly affected by the presence of an excited valley degree. 
This latter case requires microscopic computations which are not part of this paper, therefore, we assume a constant valley phase difference which we factor out. A last remark, both correction factors are unimportant in an experimental realization for the here discussed gates since this renormalization is already accounted for during the calibration process of the the qubit energy difference and the spin exchange strength.

In the presence of spin-orbit interaction, spin non-conserving tunneling is allowed which gives rise to an isotropic exchange interaction~\cite{hetenyiExchangeInteractionHolespin2020a,hendrickxFourqubitGermaniumQuantum2021}
\begin{align}
    H_\text{exchange,anisotropic}=h \boldsymbol{S}_1 \mathcal{J} \boldsymbol{S}_2,
\end{align}
where $\mathcal{J}$ is the exchange tensor. Therefore, the dynamics of the exchange interaction is no longer limited to the $\lbrace\ket{S},\ket{T_0}\rbrace$ subspace. Additionally, in the presence of SOI the quantization axis of the spins in the different quantum dots may change thus giving rise to an additional channel that can couple the $\ket{T_0}$ state with the polarized $\ket{T_+}=\ket{\uparrow,\uparrow}$ and $\ket{T_-}=\ket{\downarrow,\downarrow}$ states. Note, that this situation may also arise for electron systems with micromagnets designed to enhance SOI~\cite{miCoherentSpinPhoton2018,samkharadzeStrongSpinphotonCoupling2018}. However, most results from this paper can still be applied to platforms with small to medium strength SOI, such as planar electron qubits in silicon and hole qubits~\cite{scappucciGermaniumQuantumInformation2020}, as long as a sufficiently strong external magnetic field is applied to suppress to unwanted dynamics.

\section{Generalized rotating wave approximation}
\label{method:generalRWA}
In our framework we now apply an ``exact'' rotating wave approximation described in Ref.~\cite{zeuchExactRotatingWave2020}. Although other approaches based either on the Magnus expansion~\cite{figueiredoroqueEngineeringFastHighfidelity2021} or Floquet engineering~\cite{dengDynamicsTwolevelSystem2016} or numerical methods~\cite{safaeiOptimizedSinglequbitGates2009} allow to find control pulses for a strongly driven qubit system we find the exact RWA to be most intuitive to incorporate in non-linear situations. The effective Hamiltonian is then given by~\cite{zeuchExactRotatingWave2020}
\begin{align}
     H_\text{effective} =& \phantom{+} \nu_D\int_{0}^{\nu_D^{-1}}dt^\prime H_\text{RF}(t^\prime) 
     \nnb &+ 
     \nu_D\int_{0}^{\nu_D^{-1}}dt^\prime \dot{H}_\text{RF}(t^\prime) t^\prime 
     \nnb &+ 
     \frac{\nu_D}{2\hbar i}\int_{0}^{\nu_D^{-1}}dt^\prime\int_{0}^{t^\prime}dt^{\prime\prime} [H_\text{RF}(t^\prime),H_\text{RF}(t^{\prime\prime})] 
     \nnb &+ 
     \mathcal{O}\left(\frac{1}{\nu_D}\right)^2,
     \label{eq:ERWA}
\end{align}

where $[A,B]=AB-BA$ denotes the commutator and $\dot{A}=\frac{d}{dt} A$ is the time derivative. An important remark is, that $H_\text{RF}$ and $\dot{H}_\text{RF}$ are assumed to be either constant or periodic in the time interval $[0,\nu_{D}^{-1})$. The first line in Eq.~\eqref{eq:ERWA} is the conventional rotating wave approximation. The second line addresses corrections due to the time-dependence of signals such as envelopes of the applied signals. The last line includes the first order corrections of the rotating wave approximation such as the Bloch-Sigert shift.

\section{Approximating the integro-differential equation using a ordinary differential equation}
\label{method:IDEtoODE}
Writing out Eq.~\eqref{eq:optimized_shape_ac} we find the integro-differential equation
\begin{widetext}
\begin{align}
    \frac{\dot{\theta}_j(t)}{4 \pi } &= \frac{2 \pi}{2 \nu_{ST}^2} \left[(\Delta E_z J(t) \cos(2\pi\nu_{ST} t+\theta_j(t) )-J_0 \Delta E_z(t) \cos(2\pi\nu_{ST} t+\theta_j (t))\right] \nonumber\\
    &\times \left(\int_0^t (\Delta E_z J(s) \sin (2 \pi  \nu_{ST} s+\theta_j (s))-J_0 \Delta E_z(s) \sin (2 \pi  \nu_{ST} s+\theta_j(s))) \, ds\right)+\frac{\Delta E_z \Delta E_z(t)+J_0 J(t)-\nu_{ST}^2}{2 \nu_{ST}}.
\end{align}
\end{widetext}
The upper system is generally hard to solve, thus we approximate the upper expression using a series expansions in $\theta_j(t)$ at $\theta_j(0)$ up to first order. To arrive at an ordinary differential equation (ODE) we additionally ignore all terms under the integral which depend on $\theta_j$. A numerical check confirmed the validity assuming reasonable smooth pulse shapes.

\section{Master equation solver}
\label{method:Simulation}
For all numerical simulations performed we solve the time-dependent Schr{\"o}dinger equation 
\begin{align}
    i\hbar\frac{d}{dt} \ket{\psi(t)} = H \ket{\psi(t)}
\end{align}
and compute the unitary propagator iterative according to
\begin{align}
    U(t+\Delta t) = e^{-\frac{i}{\hbar} H(t+\Delta t)}U(t).
\end{align}
Here, $H(t+\Delta t)$ is discretized into $N$ segments of length $\Delta t$ such that $H(t)$ is constant in the time-interval $\left[t,t+\Delta t\right)$.
For the simulations of single-qubit gates in Figs.~\ref{fig:main_figure_1}~(b)-(d) we choose an external magnetic field $E_z=\unit[10]{GHz}$ and a step-size of $\Delta t=\unit[0.2]{ps}$.
The remaining simulations involving two-qubit gates in Figs.~\ref{fig:main_figure_2}~and~\ref{fig:main_figure_3} are performed in the rotating frame of the external magnetic field $E_z$ and neglecting the counter-rotating terms, the so-called rotating wave approximation (RWA). This allows us to chose the time-step $\Delta t=\unit[10]{ps}$ being sufficiently small. For each simulation we check whether the RWA holds.

In all our simulations except for ODE~\ref{fig:main_figure_3}~(a)~(green) to emulate the realistic effect of a finite band-with and filtering of the control gates we add on each input signal, amplitude and phase of barrier voltage and drive voltage, a low-pass filter with a cut-off frequency of $\unit[150]{MHz}$. In particular we use a butterworth filter of order 3 implemented via the Scipy package~\cite{virtanenSciPyFundamentalAlgorithms2020}.

For our noisy simulations we included classical fluctuations of the matrix elements of the Hamiltonian
\begin{align}
    H(t,\bo{\beta}(t)) \rightarrow H(t,\bo{\beta}(t)+\delta\bo{\beta}(t)), 
\end{align}
where $\bo{\beta}(t)$ are parameters describing the dynamics of the system and $\delta\bo{\beta}(t)$ are the fluctuations of the parameter. We note, that this treatment allows us to account for non-linear Hamiltonian interactions as the exchange interaction and also to include 'sweet spots', points of operation where the first order dynamics vanishes.
The fluctuations $\delta\beta_j(t)$ are described by its spectral density $S_{\delta\beta_j}(w)=\int_{-\infty}^\infty \left(\int_{-\infty}^\infty \delta\beta_j(t)\delta\beta_j(t-t^\prime) dt^\prime\right)e^{-iw t} dt$ which we use as input in our simulations. To compute time-traces of the fluctuation we use the Fourier filtering method~\cite{yangAchievingHighfidelitySinglequbit2019,koskiStrongPhotonCoupling2020} to generate time-correlated time traces obeying $S_{\delta\beta_j}(w)$. In this method $N$ independent Gaussian distributed numbers are generated and associated to a segment with time $\Delta t$. We then perform a discrete Fourier transformation, apply a frequency filter $\sqrt{S_{\delta\beta_j}(\Delta w)}$, and perform an inverse Fourier transformation. The resulting fluctuations are discretized in $N$ segments with time $\Delta t$ such that $\delta\beta_j(t)$ is constant in the time interval $\left[t,t+\Delta t\right)$. Note, that the same $\Delta t$ is used as above. Consequently, fluctuations which are faster than $f\st{max}=\frac{1}{\Delta t}$ are truncated. The lower frequency cut-off of the discretized noise traces is given by the length of the simulation and the number of segments. In order to provide better comparison of the dynamics of unequal length simulations and avoid an increase of simulations length we add static fluctuations $\beta\st{static}$ to compensate for the 'missing' low-frequency components. The amount of static noise component is given by 
\begin{align}
    \text{Var}(\beta\st{j,static}) = \frac{1}{\pi}\int_{2\pi f\st{min}}^{\frac{2\pi}{t\st{sim}}} S_{\delta\beta_j}(w) d\omega,
\end{align}
where $t\st{sim}$ is the length of the simulation and $f\st{min} = \unit[0.1]{Hz}$ is a lower-frequency cut-off and resembles the retuning cycle of an experiment. 

The final superoperator of the noisy process is then constructed as follows
\begin{align}
    \chi = \frac{1}{N_j} \sum_j U^\dagger_\text{j}(t_g) \otimes U_\text{j}(t_g),
\end{align}
where $U_j(t=t_g)$ is the final unitary time evolution operator for a noise realization $j$ and $N_j=5000$ is the total number of noise realizations.

\section{ of coherent \textsc{cz} gate error in Mills.~et~al.}
\label{appendix:mills_coherent}
We find the following device parameter in Ref.~\cite{millsTwoqubitSiliconQuantum2022} describing the \textsc{cz} gate; difference in qubit resonance frequency $\Delta\widetilde{ E}_z=\unit[396]{MHz}$, duration of \textsc{cz} gate $t_g=\unit[40]{ns}$ using a (smoothed) rectangular pulse shape (Tukey window with $\lambda\approx 0$). From Eq.~\ref{eq:error_rate_2Q} we know that the \textsc{cz} error probability only depends on $\Delta E_z t_g$. This invariant allows us to conveniently simulate the coherent error using $\Delta E_z=\unit[100]{MHz}$ (as for all other simulations in this manuscript) but using a modified pulse time $\widetilde{t}_g=t_g \times \Delta\widetilde{ E}_z/\Delta E_z = \unit[158.4]{ns}$. For the noisy simulations we used the same noise strength as extracted from Ref.~\cite{xueQuantumLogicSpin2022a} (as for all other simulations in this manuscript). We argue that the longer gate time (thus more noise) is compensated by the faster dephasing observed in Ref.~\cite{millsTwoqubitSiliconQuantum2022}.

\bibliography{Lit_06_22}

\end{document}